\documentclass[a4paper,12pt]{article}

\pdfoutput=1
\usepackage{amsmath}
\usepackage{amssymb}
\usepackage{amsfonts}
\usepackage{mathrsfs}
\usepackage{bbm}
\usepackage{graphicx,subfigure,booktabs}
\usepackage[numbers,sort&compress]{natbib}
\usepackage{verbatim}
\usepackage{color}
\usepackage{ulem}
\usepackage{setspace}
\usepackage{multirow}
\usepackage{url}

\usepackage[utf8]{inputenc}

\usepackage[colorlinks,
linkcolor=black,
filecolor=black,
anchorcolor=black,
urlcolor=black,
citecolor=blue,
bookmarks=false,
]{hyperref}

\usepackage[hyphenbreaks]{breakurl}

\usepackage{fancyhdr}

\numberwithin{equation}{section}

\newlength{\dinwidth}
\newlength{\dinmargin}
\allowdisplaybreaks

\begin{document}

%%%%%%%%%%%%%%%%%%%%%%%%%%%%%%%%%%%%%%%%%%%%%%%%%%
%\begin{abstract}
%abs
%\end{abstract}
%%%%%%%%%%%%%%%%%%%%%%%%%%%%%%%%%%%%%%%%%%%%%%%%%%

%\maketitle

\title{Studying the $b\rightarrow s \ell^+\ell^-$ anomalies and $(g-2)_{\mu}$ in $R$-parity violating MSSM framework with the inverse seesaw mechanism}

\author{
Min-Di Zheng\footnote{zhengmd5@mail.sysu.edu.cn}\,\,
and 
Hong-Hao Zhang\footnote{zhh98@mail.sysu.edu.cn}\\[15pt]
\small School of Physics, Sun Yat-Sen University, Guangzhou 510275, China}

\date{}
\maketitle

%%%%%%%%%%%%%%%%%%%%%%%%%%%%%%%%%%%%%%%%%%%%%%%%%%%%
\begin{abstract}
Inspired by the recent experimental results which show deviations from the standard model (SM) predictions of $b\rightarrow s \ell^+\ell^-$ transitions, we study the $R$-parity violating minimal supersymmetric standard model (RPV-MSSM) extended by the inverse seesaw mechanism. The trilinear $R$-parity violating terms, together with the chiral mixing of sneutrinos, induce the loop contributions to the $b\rightarrow s \ell^+\ell^-$ anomaly. We study the parameter space of the single-parameter scenario $C^{\rm NP}_{9,\mu}=-C^{\rm NP}_{10,\mu}=C_{\rm V}$ and the double-parameter scenario $(C_{\rm V},C_{\rm U})$,  respectively, constrained by other experimental data such as $B_s-\bar{B}_s$ mixing, $B\rightarrow X_s \gamma$ decay, the lepton flavour violating decays, etc. Both the single-parameter and the double-parameter scenario can resolve the long existing muon anomalous magnetic moment problem as well, and allow the anomalous $t\rightarrow cg$ process to reach the sensitivity at the Future Circular hadron-hadron Collider (FCC-hh).    
\end{abstract} 
%%%%%%%%%%%%%%%%%%%%%%%%%%%%%%%%%%%%%%%%%%%%%%%%%%%
\newpage

\section{Introduction}
\label{sec:intro}
In recent years, several hints of new physics (NP) beyond the SM have shown up, such as, $R_{K^{({\ast})}} = {{\mathcal B}(B \rightarrow K^{({\ast})} \mu^+ \mu^-)}/{{\mathcal B}(B \rightarrow K^{({\ast})} e^+ e^-)}$ on the transitions of $b \rightarrow s \ell^+\ell^-$ ($\ell=e,\mu$), which exhibits the very attractive anomalies. Especially, the measurement of $R_K$ by the LHCb Collaboration has just been updated with the full run {\rm II} data as $R_K=0.846^{+0.042~+0.013}_{-0.039~-0.012}$ in the $q^2$ bin $[1.1,6]$ ${\rm GeV}^2$~\cite{Aaij:2021vac}, which is much more precise than the previous one $R_K=0.846^{+0.060~+0.016}_{-0.054~-0.014}$~\cite{Aaij:2019wad}, giving rise to the discrepancy with the SM value changing from preceding $2.5\sigma$ to $3.1\sigma$. 
The recent measurements of $R_{K^{\ast}}$ by LHCb give $R_{K^\ast}=0.66^{+0.11}_{-0.07}\pm0.03$ at $[0.045,1.1]$ ${\rm GeV}^2$ bin and $R_{K^\ast}=0.69^{+0.11}_{-0.07}\pm0.05$ at $[1.1,6]$ ${\rm GeV}^2$ bin, 
showing $2.1\sigma$ deviation at low $q^2$ region and $2.5\sigma$ deviation at high region, respectively~\cite{Aaij:2017vbb}.  
The $R_{K^{(\ast)}}$ results by the Belle Collaboration~\cite{Abdesselam:2019lab,Abdesselam:2019wac} 
show the consistency with the SM predictions although the results have sizable experimental error bars. Besides, there are also other anomalies in the $b\rightarrow s \ell^+\ell^-$ transition, for instance, the angular observable $P'_5$ anomaly of $B \rightarrow K^\ast \mu^+ \mu^-$ decay persists with the new data~\cite{Aaij:2020nrf} compared with the run {\rm I} results~\cite{Aaij:2015oid,Aaij:2013qta,Khachatryan:2015isa,Aaboud:2018krd,Wehle:2016yoi,Abdesselam:2016llu}. 
All these anomalies may indicate the NP that breaks lepton 
flavour universality (LFU).

Ones know that each single anomaly above cannot be  regarded as the conclusive evidence of NP. However, 
it is interesting that nearly all these anomalies can be explained simultaneously with the four-Fermi operators in the model-independent global fit~\cite{Aebischer:2019mlg,Alok:2019ufo,Alguero:2019ptt,Ciuchini:2019usw,Arbey:2019duh,Kowalska:2019ley,Capdevila:2019tsi,Bhattacharya:2019dot,Bhutta:2020qve,Biswas:2020uaq,Alda:2020okk,Carvunis:2021jga,Geng:2021nhg,Li:2021toq,Angelescu:2021lln,Altmannshofer:2021qrr,Cornella:2021sby,Kriewald:2021hfc,Isidori:2021vtc,Alguero:2021anc,Hurth:2021nsi,Perez:2021ddi,Alda:2021ruz,Bause:2021ply}. In light of the new measurement of $R_K$~\cite{Aaij:2021vac}, there are already some new fit results updated~\cite{Alok:2019ufo,Alda:2020okk,Carvunis:2021jga,Geng:2021nhg,Li:2021toq,Angelescu:2021lln,Altmannshofer:2021qrr,Cornella:2021sby,Kriewald:2021hfc,Isidori:2021vtc,Alguero:2021anc,Hurth:2021nsi,Perez:2021ddi,Alda:2021ruz,Bause:2021ply}.
For the discussion of the global fit, the related Lagrangian of low energy effective field theory is given by
\begin{equation}
{\cal L}_{\rm eff} =  \frac{4 G_F}{\sqrt{2}} \eta_t \sum_{i}C_i{\cal O}_i + {\rm h.c.} ,
\end{equation} 
where the Cabibbo–Kobayashi–Maskawa (CKM) factor $\eta_t \equiv K_{tb} K_{ts}^{\ast}$. The main operators for the anomaly explanations are
\begin{align}
{\cal O}_9 = \frac{e^2}{16\pi^2}(\bar{s}\gamma_\mu P_{L} b)(\bar{\ell}\gamma^\mu \ell), \quad
{\cal O}_{10} = \frac{e^2}{16\pi^2}(\bar{s}\gamma_\mu P_{L} b)(\bar{\ell}\gamma^\mu \gamma_5 \ell),
\end{align}
where $P_L=(1-\gamma_5)/2$ is the left-handed (LH) chirality projector and the Wilson coefficients $C_{9(10)} = C_{9(10)}^{\rm SM} + C_{9(10)}^{\rm NP}$. 
In this work, we adopt the following unified form of fit scenarios
\begin{align}\label{eq:sce}
C^{\rm NP}_{9,\mu}=&C_{\rm V}+C_{\rm U}, 
\quad
C^{\rm NP}_{10,\mu}=-C_{\rm V}, \notag\\
C^{\rm NP}_{9,e} =&C_{\rm U}, \quad
C^{\rm NP}_{10,e} =0,
\end{align}
where ${\rm V}$ denotes the contributions only of $\mu^+\mu^-$ channel and ${\rm U}$ denotes LFU contributions. 
The first scenario, called scenario A here, requires $C_{\rm U}=0$ to realize the single-parameter scenario $C^{\rm NP}_{9,\mu}=-C^{\rm NP}_{10,\mu}$ in fact. We adopt the fit result 
$-0.46<C_{\rm V}<-0.32$ that conforms to the rare $B$-meson decays at $1\sigma$ level in Ref.~\cite{Altmannshofer:2021qrr}. Except the new $R_K$ measurement~\cite{Aaij:2021vac}, authors in Ref.~\cite{Altmannshofer:2021qrr} have also considered other series of new experimental results, such as the new angular analyses of $B^0\rightarrow K^{\ast 0}\mu^+\mu^-$~\cite{Aaij:2020nrf} and $B^\pm\rightarrow K^{\ast \pm}\mu^+\mu^-$~\cite{Aaij:2020ruw}, the updated branching ratio measurement of the $B_s\rightarrow \phi\mu^+\mu^-$~\cite{LHCb:2021zwz} that confirms the previous tension~\cite{LHCb:2015wdu} with the SM prediction, 
as well as the recent results of $B_s\rightarrow \mu^+\mu^-$ from CMS~\cite{Sirunyan:2019xdu} and LHCb~\cite{LHCb:2021awg,LHCb:2021vsc}. 
For the case of $C_{\rm U}\neq 0$, named as scenario B, we also utilize 
the fit regions in Ref.~\cite{Altmannshofer:2021qrr} with the best fit point $(C_{\rm V},C_{\rm U})\approx(-0.34,-0.32)$.
 
After these results of the model-independent analyses are obtained, imperative works are to find the concrete NP models which can conform to them. Both the scenario A and B have been implemented in RPV-MSSM~\cite{Deshpand:2016cpw,Das:2017kfo,Earl:2018snx,Hu:2019ahp,Hu:2020yvs,Altmannshofer:2020axr}. 
When masses of sneutrinos/sd-quarks are sufficiently heavy or there is a cancellation in the penguin contribution~\cite{Hu:2019ahp}, the scenario B turns into the scenario A.

More than RPV-MSSM, the seesaw mechanism~\cite{Minkowski:1977sc,Sawada:1979dis,GellMann:1980vs,Mohapatra:1979ia,Schechter:1980gr,Schechter:1981cv,Lazarides:1980nt} is also researched for the explanation of 
$b \rightarrow s \ell^+\ell^-$ disparities in the supersymmetric (SUSY) models~\cite{Khalil:2017mvb}, two-Higgs doublet~\cite{Li:2018rax,DelleRose:2019ukt} and other frameworks (see Refs.~\cite{Hue:2016nya,Bhatia:2017tgo,Ko:2017quv,Dinh:2017smk,Chiang:2017hlj,Antusch:2017tud,King:2018fcg,Heeck:2018ntp,Borah:2020swo,Cen:2021iwv,Alvarado:2021nxy,Dutta:2021afo,Greljo:2021xmg,Greljo:2021npi,Bhatia:2021eco} e.g.).
The seesaw mechanism is one of the most attractive methods to generate the neutrino masses in accord with the conclusive evidence of neutrino oscillations~\cite{Esteban:2020cvm}. As one type of seesaw mechanisms, the inverse seesaw~\cite{Mohapatra:1986aw,Mohapatra:1986bd} can give a $\mathcal{O}(1)$ neutrino Yukawa coupling $Y_\nu$. 
The relative large $Y_\nu$ implicates that the admixture between LH neutrino superfields and right-handed (RH) or extra singlet ones is not negligible.
Therefore, it is meaningful to study the chiral mixings of (s)neutrinos in MSSM framework extended by both inverse seesaw mechanisms and the tree-level trilinear RPV terms, and then all chiral parts of (s)neutrinos more than LH ones can interact with (s)quarks. This new combination is naturally reasonable~\cite{Escudero:2008jg,Fidalgo:2011tm,Lopez-Fogliani:2017qzj} and has never been studied in the $b \rightarrow s \ell^+\ell^-$ anomalies to our knowledge.

The recent experimental results of $R(D^{(\ast)})={\cal B}(B \to D^{(\ast)} \tau \nu) / {\cal B}(B \to D^{(\ast)} \ell \nu)$ in charged current $b\rightarrow c \tau \nu$ from BaBar~\cite{Lees:2012xj,Lees:2013uzd}, Belle~\cite{Huschle:2015rga,Hirose:2016wfn,Hirose:2017dxl,Belle:2019rba} and LHCb~\cite{Aaij:2015yra,Aaij:2017uff,Aaij:2017deq} have been averaged by the Heavy Flavor Averaging Group~\cite{Amhis:2019ckw}, and also show the tension with the average of SM predictions~\cite{Bigi:2016mdz,Bernlochner:2017jka,Bigi:2017jbd,Jaiswal:2017rve} and the recent new SM results~\cite{Gambino:2019sif,Bordone:2019vic,Jaiswal:2020wer,Cheung:2020sbq,Choi:2021mni,Iguro:2020cpg}. 
While the new measurements of $R(D^{(\ast)})$ from Belle using the semileptonic tagging method, such as the latest results with the data sample of $772\times10^6$ $B\bar{B}$ pairs, are already in agreement with SM predictions well, and Belle combined results are consistent with SM predictions within $1.6\sigma$~\cite{Hara:2020nom}. Given this, in this work, we do not investigate $R(D^{(\ast)})$ for the moment.

The clues of LFU violation exist not only in $B$-meson decays but also in other processes, such as, the muon anomalous magnetic moment problem which has existed for several decades. The measurement of $a_\mu=(g-2)_{\mu}$ by Fermilab~\cite{Abi:2021gix,Albahri:2021ixb,Albahri:2021kmg} presents the $3.3\sigma$ deviation greater than the SM prediction~\cite{Aoyama:2020ynm}\footnote{A recent calculation of the hadronic vacuum polarization (HVP)~\cite{Borsanyi:2020mff} weakens the discrepancy between the experiment and SM prediction of $a_\mu$ while it shows the tension with the R-ratio determinations~\cite{Colangelo:2018mtw,Davier:2019can,Keshavarzi:2019abf,Hoferichter:2019mqg}. And even the large HVP contribution can account for the measurement of $a_\mu$, there exists the tension within the electroweak (EW) fit~\cite{Crivellin:2020zul,Keshavarzi:2020bfy,deRafael:2020uif,Malaescu:2020zuc}. 
We do not consider this HVP result here but consider the preceding review of various SM results~\cite{Aoyama:2020ynm}.}, and agrees with the previous Brookhaven E821 experiment~\cite{Bennett:2006fi}. The combined deviation average of the two experiments, $\Delta a_\mu= a_\mu^{\rm{exp}}-a_\mu^{\rm{SM}}=(2.51\pm 0.59)\times 10^{-9}$ shows the increased tension at the significant $4.2\sigma$ level and this is a growing motivation of NP. 
For the electron anomalous magnetic moment, there is a negative $2.4\sigma$ discrepancy between the measurement~\cite{Hanneke:2010au} and the SM prediction~\cite{Aoyama:2017uqe}, $\Delta a_e= a_e^{\rm{exp}}-a_e^{\rm{SM}}=(-8.7\pm 3.6)\times 10^{-13}$, due to the new  measurement of the fine structure constant in Ref.~\cite{Parker:2018vye}\footnote{Another new measurement of the fine structure constant~\cite{Morel:2020dww} differs by more than $5\sigma$ to the previous one~\cite{Parker:2018vye} and affects the  deviation $\Delta a_e$ to positive $1.6\sigma$ level~\cite{Gerardin:2020gpp}. The NP hint search in $a_e$ still expects more experimental researches and we focus on $a_\mu$ anomaly explanations in this work.}.
There are plentiful articles discussing the $(g-2)_\mu$ problem in the SUSY framework (e.g., see Refs.~\cite{Hisano:2001qz,Kim:2001se,Martin:2001st,Stockinger:2006zn,Belyaev:2016oxy,Yamaguchi:2016oqz,Choudhury:2017fuu,Choudhury:2017acn,Altin:2017sxx,Kotlarski:2019muo,Endo:2019bcj,Badziak:2019gaf,Kpatcha:2019pve,Altmannshofer:2020axr,Heinemeyer:2021opc,Nagai:2020xbq,Yin:2020afe,Yang:2020bmh,Han:2020exx,Yin:2021yqy,Yin:2021mls,Cao:2019evo,Cao:2021lmj,Cao:2021tuh,Zhang:2021gun,Endo:2021zal,Ahmed:2021htr,Baum:2021qzx,Abdughani:2021pdc,Ibe:2021cvf,VanBeekveld:2021tgn,Cox:2021gqq,Han:2021ify,Gu:2021mjd,Wang:2021bcx,Aboubrahim:2021xfi,Yang:2021duj,Athron:2021iuf,Altmannshofer:2021hfu,Zhang:2021nzv,Li:2021pnt,BhupalDev:2021ipu,Kim:2021suj,Ellis:2021zmg,Zhao:2021eaa,Frank:2021nkq,Li:2021koa,Aboubrahim:2021ily,Nakai:2021mha,Li:2021xmw,Ke:2021kgy}). In this work, we will investigate whether the parameter space for the explanation of $b\rightarrow s\ell^+\ell^-$ anomalies can be in accord with the deviation $\Delta a_\mu$, and then discuss the NP effects on $a_e$. 

Our paper is organized as follows. At first, the new model in this work is introduced in section~\ref{sec:model}. Then in section~\ref{sec:bsll}, the whole one-loop contributions to the $b\rightarrow s \ell^+\ell^-$ transition in this model are scrutinized and we emphasize the main contributions to explain the $b\rightarrow s \ell^+\ell^-$ anomaly in our parameter scheme. We discuss NP contributions to $(g-2)_{\ell}$ and other related constraints in section~\ref{sec:amu&constraints} followed by the numerical results and discussions in section~\ref{sec:num}. Our conclusions are presented in section~\ref{sec:conclusion}.

%%%%%%%%%%%%%%%%%%%%%%%%%%%%%%%%%%%%%%%%%%%
\section{The Model}\label{sec:model}
The model considered in this work is RPV-MSSM with inverse seesaw mechanisms called RPV-MSSMIS here and the superpotential is expressed by
\begin{align}\label{eq:MSSMIS-RPV}
{\cal W} = {\cal W}_{\rm MSSM} 
+ Y_\nu^{ij} \hat R_i \hat L_j \hat H_u + M_R^{ij} \hat R_i \hat S_j + \frac{1}{2} \mu_S^{ij} \hat S_i \hat S_j
+ \lambda'_{ijk} \hat L_i \hat Q_j \hat D_k,
\end{align}
where the generation indices $i,j,k=1,2,3$ while the colour ones are omitted. All the repeated indices are defaulted to be summed over unless otherwise stated. Here the superpotential of MSSM~\cite{Rosiek:1989rs,Rosiek:1995kg} is expressed as
\begin{align}
{\cal W}_{\rm MSSM}=\mu \hat H_u \hat H_d
+ Y_u^{ij} \hat U_i \hat Q_j \hat H_u
- Y_d^{ij} \hat D_i \hat Q_j \hat H_d
- Y_e^{ij} \hat E_i \hat L_j \hat H_d.
\end{align}
In RPV-MSSMIS, MSSM superfields are extended by three generations of pairs of SM singlet superfields, $\hat R_i$ and $\hat S_i$. The neutral parts of two Higgs doublet superfields $\hat H_u=(\hat H^+_u,\hat H^0_u)^T$ and $\hat H_d=(\hat H^0_d,\hat H^-_d)^T$ acquire the non-zero vacuum expectation values, $\langle \hat H^0_u \rangle=v_u$ and $\langle \hat H^0_d \rangle=v_d$ respectively, leading to the mixing angle $\beta=\tan^{-1}(v_u/v_d)$. The tree-level trilinear RPV coupling $\lambda'_{ijk} \hat L_i \hat Q_j \hat D_k$ can be added for $\hat L_i$ sharing the same SM quantum number with $\hat H_d$. It is needed to point out that the RPV superpotential terms like $\lambda'_{ijk} \hat L_i \hat Q_j \hat D_k$, $\lambda_{ijk} \hat L_i \hat L_j \hat E_k$, $\lambda''_{ijk} \hat U_i \hat D_j \hat D_k$ as well as $\mu_i \hat L_i \hat H_u$ are all in principle allowed for the SM gauge invariance if there are no extra symmetries. Here we only consider the term $\lambda'_{ijk} \hat L_i \hat Q_j \hat D_k$ connecting the quark sector with lepton sector, without the pure-quark term $\lambda''_{ijk} \hat U_i \hat D_j \hat D_k$ or the purely lepton interaction $\lambda_{ijk} \hat L_i \hat L_j \hat E_k$, because of the attempt to avoid the probable  disastrously rapid proton decay~\cite{Weinberg:1981wj,Sakai:1981pk} when there are nonzero parameters $\lambda'$ and $\lambda''$ simultaneously and
the strong collider constraints on the lightest sneutrino mass when the $\lambda'$ and $\lambda$ both exist~\cite{Aaltonen:2010fv,Abazov:2010km,Aad:2015pfa,Khachatryan:2016ovq,Aaboud:2016hmk,Aaboud:2018jff,Sirunyan:2018zhy,Le:2019bbt,ATLAS:2018rns,ATLAS:2018mrn,ATLAS:2021yyr}. The bilinear term $\mu_i \hat L_i \hat H_u$ is also allowed but we exclude it in order to avoid the extra contributions to neutrino masses~\cite{Hirsch:2000ef}. 
With the scalar components of Higgs doublet superfields denoted by $H_u$ and $H_d$, and squarks and sleptons denoted by ``$\tilde{\ }$'', the soft supersymmetry breaking Lagrangian is given by
\begin{align}\label{eq:softL}
-{\cal L}^{\rm soft}=&-{\cal L}^{\rm soft}_{\rm MSSM}
+(m^2_{\tilde{R}})_{ij} \tilde{R}^\ast_i\tilde{R}_j
+(m^2_{\tilde{S}})_{ij} \tilde{S}^\ast_i\tilde{S}_j \notag \\
&+(A_\nu Y_{\nu})_{ij} \tilde{R}^\ast_i\tilde{L}_jH_u
+B_{M_R}^{ij} \tilde{R}^\ast_i\tilde{S}_j
+\frac{1}{2} B_{\mu_S}^{ij} \tilde{S}_i\tilde{S}_j,
\end{align}
where ${\cal L}^{\rm soft}_{\rm MSSM}$ corresponds to MSSM part~\cite{Rosiek:1989rs,Rosiek:1995kg}. It should be mentioned that MSSM and singlet neutrino sectors are all at low scale (around $1$ TeV) in this work, so some terms in the most general superpotential and soft breaking Lagrangian are already or will be eliminated ad hoc for the phenomenological consideration.

As to the three terms following ${\cal W}_{\rm MSSM}$ in Eq.~\eqref{eq:MSSMIS-RPV} which give the neutrino mass spectrum at tree level, the $9\times 9$ mass matrix of neutrino in the $(\nu, R, S)$ basis is given by
\begin{align}\label{eq:mnu}
{\cal M}_{\nu} = \left( 
\begin{array}{ccc}
0 &m_D^{T}  &0\\ 
m_D  &0 &M_R\\ 
0 &M_{R}^{T} &\mu_S\end{array} 
\right), 
\end{align}
in which $m_D = \frac{1}{ \sqrt{2}} v_u Y_\nu^T$. And the $\mu_S$ parameter can be obtained by
\begin{align}\label{eq:musterm}
\mu_S= (m_D^{T})^{-1} M_R U_{\text{PMNS}} m^{\text{diag}}_{\nu} U_{\text{PMNS}}^T M_R^T m_D^{-1},
\end{align}
when $\mu_S\ll m_D < M_R$.
The diagonalized neutrino mass ${\cal M}^{\text{diag}}_{\nu}$ in physics basis containing the three-light-generation part $m^{\text{diag}}_{\nu}$ in Eq.~\eqref{eq:musterm}, is given by ${\cal M}^{\text{diag}}_{\nu}={\cal V} {\cal M}_{\nu} {\cal V}^T$. Here embedded in the whole $9 \times 9$ mixing matrix ${\cal V}^T$, the $3 \times 3$ light-generation sector ${\cal V}^T_{3 \times 3}$ should approximate the PMNS matrix $U^{\rm PMNS}$~\cite{Esteban:2018azc,Esteban:2020cvm}.

Then we turn to the sneutrino mass square matrix in the $(\tilde{\nu}^{\cal I(R)}_{L},\tilde{R}^{\cal I(R)},\tilde{S}^{\cal I(R)})$ basis, which is expressed as
\begin{align}\label{eq:mSnu}
{\cal M}_{\tilde{\nu}^{\cal I(R)}}^2  = 
\left(\begin{array}{ccc} 
m^2_{\tilde{L}'} & (A_\nu -\mu\cot\beta) m_D^T & m_D^T M_R \\
(A_\nu -\mu\cot\beta) m_D & m^2_{\tilde{R}}+M_RM_R^{T}+m_Dm_D^{T} & 
       \pm M_R\mu_S + B_{M_R} \\
M_R^T m_D & \pm \mu_S M_R^{T} + B_{M_R}^T 
& m^2_{\tilde S}+ \mu_S^2+M_R^TM_R \pm B_{\mu_S}
\end{array}\right),
\end{align}
where the ``$\pm$'' above expresses the $CP$-even and $CP$-odd, and also $CP$-odd is denoted by ${\cal I}$ and $CP$-even is denoted by ${\cal R}$. The soft mass $m^2_{\tilde{L}'}=m^2_{\tilde{L}}+\frac{1}{2} m^2_Z \cos 2\beta+m_Dm_D^T$ can be regarded as one whole input where $m^2_{\tilde{L}}$ is the soft mass square of $\tilde{L}$ in ${\cal L}^{\rm soft}_{\rm MSSM}$. The $CP$-even and $CP$-odd masses can be nearly the same for tiny $\mu_S$ and relatively small $B_{\mu_S}$~\cite{BhupalDev:2012ru}. Besides, 
the value of $m^2_{\tilde S}$ is set to be zero here. Thus, the approximate form is provided as~\cite{Chang:2017qgi}
\begin{align}\label{eq:mSnuNum}
{\cal M}_{\tilde{\nu}^{\cal I(R)}}^2  \approx
\left(\begin{array}{ccc} 
m^2_{\tilde{L}'} & (A_\nu -\mu\cot\beta) m_D^T & m_D^T M_R \\
(A_\nu -\mu\cot\beta) m_D & m^2_{\tilde{R}}+M_RM_R^{T}+m_Dm_D^{T} & B_{M_R} \\
M_R^T m_D & B_{M_R}^T 
& M_R^T M_R
\end{array}\right).
\end{align}
In the following we adopt this particular structure and then the mixing matrices $\tilde{\cal V}^{\cal I(R)}$, which diagonalize sneutrino mass square matrices by $\tilde{\cal V}^{\cal I(R)} {\cal M}_{\tilde{\nu}^{\cal I(R)}}^2 \tilde{\cal V}^{\cal I(R) \dagger}={({\cal M}_{\tilde{\nu}^{\cal I(R)}}^2)}^{\text{diag}}$, are also the same whether $CP$ even or odd. All the $\tilde{\cal V}^{\cal R}$ and the physical mass $m_{\tilde{\nu}^{\cal R}}$ can be expressed as $\tilde{\cal V}^{\cal I}$ and $m_{\tilde{\nu}^{\cal I}}$, respectively in the rest of the paper. With respect to charged sleptons, the LH sector element is given by $m^2_{\tilde{L}'}+m_l^2-m_D m^T_D-m^2_W \cos2\beta$.

Afterwards we discuss the last term of the superpotential. For the superpotential term $\lambda'_{ijk} \hat L_i \hat Q_j \hat D_k$, the corresponding Lagrangian in the flavour basis is obtained as
\begin{align}\label{eq:RPVlagflav}
{\cal L}_{\text{LQD}} =& \lambda'_{ijk}\big(\tilde{\nu}_{Li} \bar{d}_{Rk} d_{Lj} + \tilde{d}_{Lj} \bar{d}_{Rk} \nu_{Li} + \tilde{d}_{Rk}^\ast \bar{\nu}_{Li}^c d_{Lj} \notag \\
&- \tilde{l}_{Li} \bar{d}_{Rk} u_{Lj} - \tilde{u}_{Lj} \bar{d}_{Rk} l_{Li} - \tilde{d}_{Rk}^\ast \bar{l}_{Li}^c u_{Lj}\big) + {\rm h.c.},
\end{align}
where ``$c$'' indicates the charge conjugated fermions.
Then in the context of mass eigenstates for the down quarks and charged leptons, the Lagrangian above with other fields $\tilde{\nu}_{L}$, $\nu_{L}$ and $u_{L}$ (aligned with $\tilde{u}_L$) rotated to mass eigenstates by mixing matrices $\tilde{\cal V}^{\cal I(R)}$, ${\cal V}$ and $K$ respectively, is given by
\begin{align}\label{eq:RPVlagphys}
{\cal L'}_{\text{LQD}} =& \lambda'^{\cal I(R)}_{vjk} \tilde{\nu}_{v} \bar{d}_{Rk} d_{Lj} 
+\lambda'^{\cal N}_{vjk} \big(\tilde{d}_{Lj} \bar{d}_{Rk} \nu_{v} + \tilde{d}_{Rk}^\ast \bar{\nu}_{v}^c d_{Lj} \big)  \notag \\
&-\tilde{\lambda}'_{ilk} \big(\tilde{l}_{Li} \bar{d}_{Rk} u_{Ll} + \tilde{u}_{Ll} \bar{d}_{Rk} l_{Li} + \tilde{d}_{Rk}^\ast \bar{l}_{Li}^c u_{Ll}\big) + {\rm h.c.},
\end{align}
where all the fields are represented by the mass eigenstates. 
Concretely, $\nu_{v}$ and $\tilde{\nu}_{v}$ are in the mass eigenstate with the index $v=1,2,\dots 9$ and the three neo-$\lambda'$ terms are deduced as $\lambda'^{\cal I(R)}_{vjk}=\lambda'_{ijk} \tilde{\cal V}^{\cal I(R) \ast}_{vi}$, $\lambda'^{\cal N}_{vjk}=\lambda'_{ijk} {\cal V}^{\ast}_{vi}$ and 
$\tilde{\lambda}'_{ilk}=\lambda'_{ijk} K^{\ast}_{lj}$. In the following, we call the diagrams including these $\lambda'$ couplings by ``$\lambda'$ diagrams'', otherwise by ``non-$\lambda'$ diagrams''.

By the end of this section, we should mention the chargino and neutralino mass matrices in the MSSM sector of this model. The chargino mass matrix is~\cite{Rosiek:1995kg}
\begin{align}
{\cal M}_{\chi^\pm}=\left(
\begin{array}{cc}
 M_2 & \sqrt{2} m_w \sin \beta \\
 \sqrt{2} m_w \cos \beta & \mu  \\
\end{array}
\right),
\end{align}
where the parameter $M_2$ is the wino mass and $\mu$ is related to the Higgsino mass. The mixing matrices $V$ and $U$ diagonalize ${\cal M}_{\chi^\pm}$ by $U^{\ast} {\cal M}_{\chi^\pm} V^{\dagger}= {\cal M}_{\chi^\pm}^{\text{diag}}$. As to the neutralino mass matrix ${\cal M}_{\chi^0}$  in the basis $(\tilde{B}$, $\tilde{W}^3$, $\tilde{H}_d^0$, $\tilde{H}_u^0)^T$~\cite{Rosiek:1995kg},
\begin{equation}\label{eq:Mchi0}
{\cal M}_{\chi^0}=\left( 
\begin{array}{cccc}
M_1 &0 &-\frac{e v_d}{2 \cos\theta_W} & \frac{e v_u}{{2} \cos\theta_W}    \\
0 &M_2   & \frac{e v_d}{2 \sin\theta_W} & -\frac{e v_u}{2 \sin\theta_W}  \\
-\frac{e v_d}{2 \cos\theta_W} & \frac{e v_d}{2 \sin\theta_W} & 0 & - \mu \\
\frac{e v_u}{{2} \cos\theta_W} & -\frac{e v_u}{2 \sin\theta_W} & - \mu & 0 
\end{array}
\right),
\end{equation}
with $M_1$ being the bino mass and $\theta_W$ being the weak mixing angle, it is diagonalized by $N {\cal M}_{\chi^0} N^T={\cal M}_{\chi^0}^{\text{diag}}$.

\section{$b \rightarrow s \ell^+\ell^-$ process in RPV-MSSMIS}\label{sec:bsll}
In RPV-MSSMIS, the tree-level diagram of the $b \rightarrow s \ell^+\ell^-$ process which exchanges $\tilde{u}_L$ makes the RH-quark-vector-current contribution
\begin{align} 
C_{9,\mu}'=-C_{10,\mu}'=-\frac{\sqrt{2}\pi^2}{G_F\eta_te^2}\frac{\tilde{\lambda}'_{2i2}\tilde{\lambda}'^{\ast}_{2i3}}{m^2_{\tilde{u}_{Li}}},
\end{align}
where the related operator ${\cal O}_{9(10)}'$ is given by $P_L$ changed into $P_R$ in ${\cal O}_{9(10)}$. 
This contribution is unwanted to explain $b \rightarrow s \ell^+\ell^-$ anomalies. Besides, the box diagrams of the $b \rightarrow s \ell^+\ell^-$ process also give such RH-quark-vector-current contributions which engage the sneutrino $\tilde{\nu}_{v^{(\prime)}}$ and LH slepton $\tilde{l}_{L\ell}$ with the coupling factors $\lambda'^{\cal I}_{vi2} \lambda'^{\cal I \ast}_{v'i3}$ and $\tilde{\lambda}'_{\ell i2}\tilde{\lambda}'^{\ast}_{\ell i3}$, respectively (see Eq.~\eqref{eq:chaWcsRH} and Eq.~\eqref{eq:neuWcs}). 
We concentrate on the loop effects of sneutrinos which are not expected heavy decoupled, and furthermore, 
the NP particles engaged in $b \rightarrow s \ell^+\ell^-$ and other related processes such as $B\rightarrow X_s \gamma$, $B_s-\bar{B}_s$ mixing, etc., are expected to have masses as sub-TeV or TeV scale, except for the heavy decoupled particles. Thus, we set the $\lambda'$ couplings taken at $0.5$~TeV scale called the $\mu_{\text{NP}}$ scale and assume $\lambda'_{ijk}$ non-negligible with the single value $k$ at this scale. It is called the single value $k$ assumption in this work, and the index $k$ is not be summed over in equations from here on. 
This kind of assumption is also taken in recent works for the similar  phenomenological consideration, such as Refs.~\cite{Deshpand:2016cpw,Das:2017kfo,Earl:2018snx,Hu:2019ahp,Hu:2020yvs}. Authors of Ref.~\cite{Earl:2018snx} further assume that $\lambda'_{ij1}=\lambda'_{ij2}=0$ which are adopted in Refs.~\cite{Hu:2019ahp,Hu:2020yvs} considering the bounds of $\tau\rightarrow \mu \rho^0$ and $\tau\rightarrow \mu \phi$ decays, while these constraints can also be negligible by setting sufficiently heavy $m_{\tilde{u}_{Lj}}$.

\begin{figure}[htbp]
	\centering
	\includegraphics[width=0.9\textwidth]{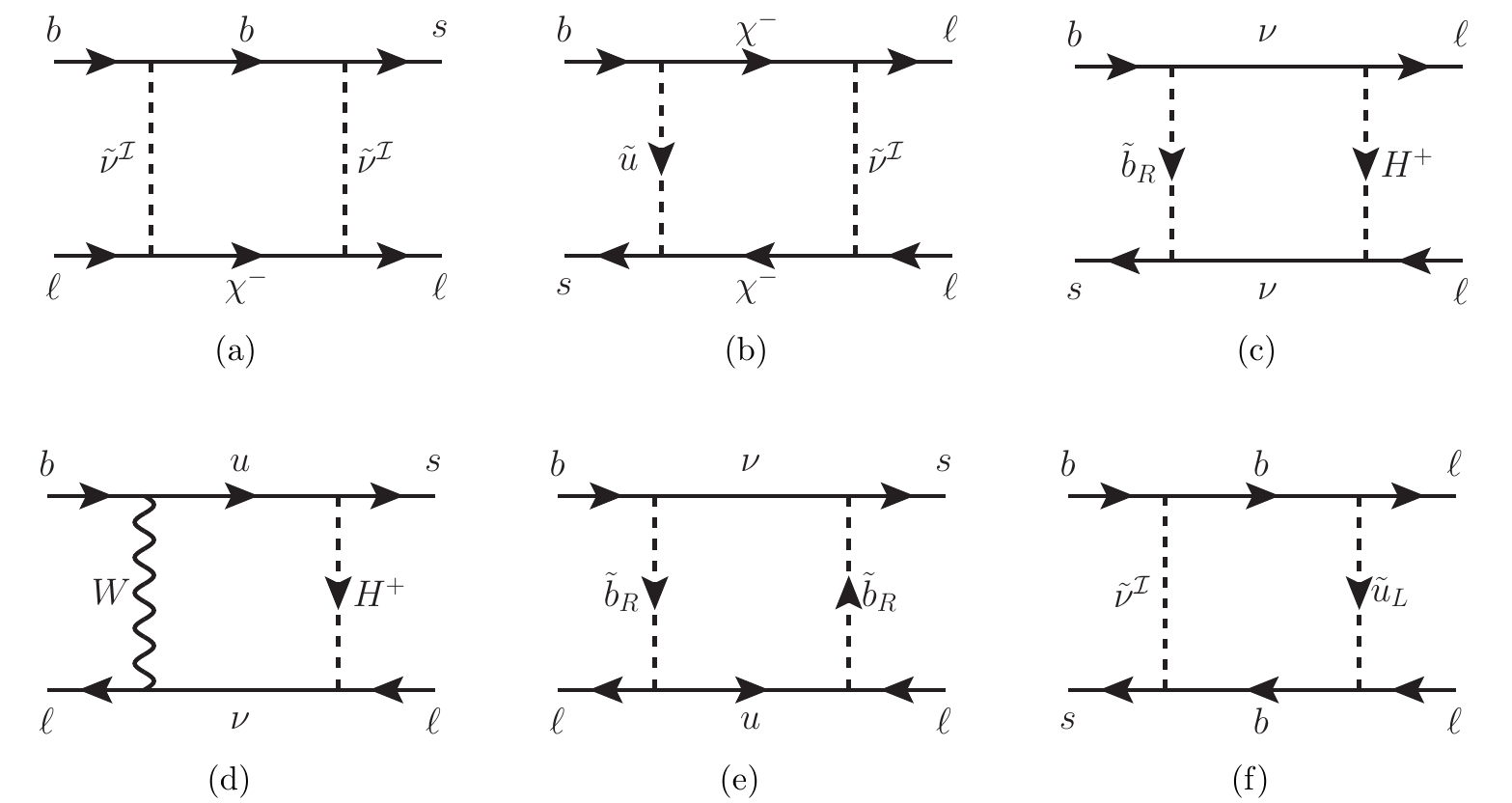}
	\caption{Box diagrams for $b \rightarrow s \ell^+\ell^-$ process in our parameter scheme. Figures \ref{fig:boxdiagram}(a) and  \ref{fig:boxdiagram}(b) show an example of $\lambda'$ diagrams and non-$\lambda'$ diagrams within chargino boxes, respectively. Figures \ref{fig:boxdiagram}(c) and \ref{fig:boxdiagram}(d) show an example of $\lambda'$ diagrams and non-$\lambda'$ diagrams within $W/H^\pm$ box, respectively. Figures \ref{fig:boxdiagram}(e) and \ref{fig:boxdiagram}(f) show the three $4\lambda'$ box diagrams with the $\tilde{\nu}^{\cal R}$-engaged diagram omitted.}
	\label{fig:boxdiagram}
\end{figure}

We scrutinize all the one-loop Feynman diagrams of $b \rightarrow s \ell^+\ell^-$ process in RPV-MSSMIS under the single value $k$ assumption. For the box diagrams, there are eleven chargino box diagrams including nine $\lambda'$ diagrams (figure~\ref{fig:boxdiagram}a) and two non-$\lambda'$ diagrams (figure~\ref{fig:boxdiagram}b), fourteen $W$ with $W$ Goldstones or charged Higgs  box (called $W/H^\pm$ box here) diagrams including ten $\lambda'$ ones (figure~\ref{fig:boxdiagram}c) and four non-$\lambda'$ ones (figure~\ref{fig:boxdiagram}d), and three $4\lambda'$ box diagrams (figure~\ref{fig:boxdiagram}e and f). The whole contributions of box diagrams are listed in the appendix (see formulas of the Passarino–Veltman functions $D_2$ and $D_0$ in Ref.~\cite{Hu:2020yvs}, where $D_{2(0)}[m^2_1, m^2_2, m^2_3, m^2_4]$ is denoted as $D_{2(0)}[m_1, m_2, m_3, m_4]$ in this paper, respectively). The Wilson coefficients in the appendix are given at $\mu_{\text{NP}}$, and if we consider the single value $k$ assumption, only the LH-quark-vector-current contributions, $C^{\rm NP}_{9(10)}(\mu_{\rm NP})$ are existent and all the RH-quark-vector-current ones, $C'^{\rm NP}_{9(10)}(\mu_{\rm NP})$ vanish. Then the Wilson coefficients run down to the scale of $\mu_b=m_b$ under QCD renormalization. One can find that $C^{\rm NP}_{9(10)}(\mu_b)\approx C^{\rm NP}_{9(10)}(\mu_{\rm NP})$~\cite{Descotes-Genon:2011nqe} and $C'^{\rm NP}_{9(10)}(\mu_b)$ vanishes due to the approximate conservation of (axial-)vector currents. Thus we can constrain the model parameters related to $C_{9(10)}(\mu_b)$ using the global fit results introduced in section~\ref{sec:intro}. 

From the appendix, one can see that $\tilde{l}_i$ and $\tilde{d}_{Lj}$ in the box diagrams can be forbidden under the assumption of a single value $k$. In the following we further assume that $m_{\tilde{d}_{Rk}}$ is sufficiently heavy to focus on the contributions of sneutrinos as the bridge between the trilinear RPV term and the inverse seesaw mechanism. Therefore, contributions including ${\tilde{d}_{Rk}}$ are negligible and can be removed. Besides, we also set $m_{\tilde{u}_{Lj}}$ adequately heavy\footnote{In this work, $m_{\tilde{d}_{Rk}}$ and $m_{\tilde{u}_{Lj}}$ are set as around $10$ TeV.}. Because the LFU violating contributions mainly from $\mu^+\mu^-$ channel are expected to explain $b \rightarrow s \ell^+\ell^-$ anomalies in both scenario A and B, we will set that ${\cal M}_{\tilde{\nu}^{\cal I(R)}}^2$ has no flavour mixing and the electron-flavour elements with both LH and RH chirality are sufficiently heavy.  
Then nearly all box contributions to $b\rightarrow se^+e^-$ transition and some box contributions to $b\rightarrow s\mu^+\mu^-$ transition can be eliminated and afterwards we show which contributions remain.

Firstly among these non-negligible chargino box diagrams, the non-$\lambda'$ diagram with RH sneutrinos previously discussed in Ref.~\cite{Khalil:2017mvb} is recalculated by us. We find the Wilson coefficient $C_9^{\text{NP}}$ from this diagram equals to $-C_{10}^{\text{NP}}$, which is different from the condition that $C_9^{\text{NP}}=C_{10}^{\text{NP}}$ in Ref.~\cite{Khalil:2017mvb}. The related $C_9^{\text{NP}}$, namely as  $C^{\chi^\pm(1)}_{\rm V}$ in this paper, is given by
\begin{align}\label{eq:chaWcs1}
C^{\chi^\pm(1)}_{\rm V}=&-\frac{\sqrt{2}\pi^2 i}{2G_F \eta_t e^2} 
y^2_{u_i} K_{i3}K^{\ast}_{i2} V^{\ast}_{m2}V_{n2}
(g_2V_{m1}\tilde{\cal V}^{\cal I}_{v2}-
V_{m2}Y^{\cal I}_{2v}) \notag\\
&(g_2V^{\ast}_{n1}\tilde{\cal V}^{\cal I}_{v2}-
V^{\ast}_{n2}Y^{\cal I}_{2v}) 
D_2[m_{\tilde{\nu}^{\cal I}_{v}},m_{\chi^\pm_m},m_{\chi^\pm_n},m_{\tilde{u}_{Ri}}],
\end{align}
where the Yukawa couplings $y_{u_i}={\sqrt{2}m_{u_i}}/{v_u}$ and $Y^{\cal I}_{\ell v}={(Y_\nu)}_{j\ell} \tilde{\cal V}^{\cal I \ast}_{v(j+3)}$. 
This formula can be seen in the second term of Eq~\eqref{eq:chaWcs}.

Then we show the $\lambda'$ within chargino box diagram containing the RPV interactions between singlet sneutrinos and quarks. The contribution is given by
\begin{align}\label{eq:chaWcs2}
C^{\chi^\pm(2)}_{\rm V}=&\frac{\sqrt{2}\pi^2 i}{2G_F \eta_t e^2}
\lambda'^{\cal I}_{v3k} \lambda'^{\cal I \ast}_{v'2k} (g_2 V^{\ast}_{m1} \tilde{\cal V}^{\cal I}_{v2}-V^{\ast}_{m2} Y^{\cal I}_{2v}) \notag\\
&(g_2 V_{m1} \tilde{\cal V}^{\cal I}_{v'2}-V_{m2} Y^{\cal I}_{2v'}) D_2[m_{\tilde{\nu}^{\cal I}_{v}},m_{\tilde{\nu}^{\cal I}_{v'}},m_{\chi^\pm_m},m_{d_k}], 
\end{align}
which appears in the third term of Eq~\eqref{eq:chaWcs}.

The non-ignorable $W/H^\pm$ box contributions in Eq.~\eqref{eq:WHWcs} are
\begin{align}\label{eq:WHWcs1}
C^{W/H^\pm(1)}_{9,\ell}=&-C^{W/H^\pm(1)}_{10,\ell}=-\frac{\sqrt{2}\pi^2 i}{2G_F \eta_t e^2} \Bigl(
y^2_{u_i} K_{i3}K^{\ast}_{i2} Z^2_{H_{h2}}Z^2_{H_{h'2}} |Y^{\cal N}_{\ell v}|^2 D_2[m_{\nu_v},m_{u_i},m_{H_h},m_{H_{h'}}] \notag\\
&-4g^2_2 m_{u_i}y_{u_i} m_{\nu_v} K_{i3}K^{\ast}_{i2} Z^2_{H_{h2}} {\text{Re}}({\cal V}_{v\ell} Y^{\cal N \ast}_{\ell v}) D_0[m_{\nu_v},m_{u_i},m_W,m_{H_{h}}] \notag\\
&+5g^4_2 K_{i3}K^{\ast}_{i2} |{\cal V}_{v\ell}|^2 D_2[m_{\nu_v},m_{u_i},m_W,m_W]
\Bigr),
\end{align}
where the mixing matrix elements $Z_{H_{12}}=-\sin \beta$, $Z_{H_{22}}=-\cos\beta$ with Goldstone mass $m_{H_1}=m_W$ and charged Higgs mass $m_{H_2}=m_{H^\pm}$ and $Y^{\cal N}_{\ell v}={(Y_\nu)}_{j\ell} {\cal V}^{\ast}_{v(j+3)}$.  
It is obvious that these $W/H^\pm$ box contributions include SM effects, which cannot be separated naively from NP effects because of the generation and the chiral mixing of massive neutrinos. In addition, these contributions still contain both the $\mu^+\mu^-$ channel sector and $e^+e^-$ channel sector.
We will further investigate these contributions in detail at section~\ref{sec:num}.

\begin{figure}[htbp]
	\centering
	\includegraphics[width=1\textwidth]{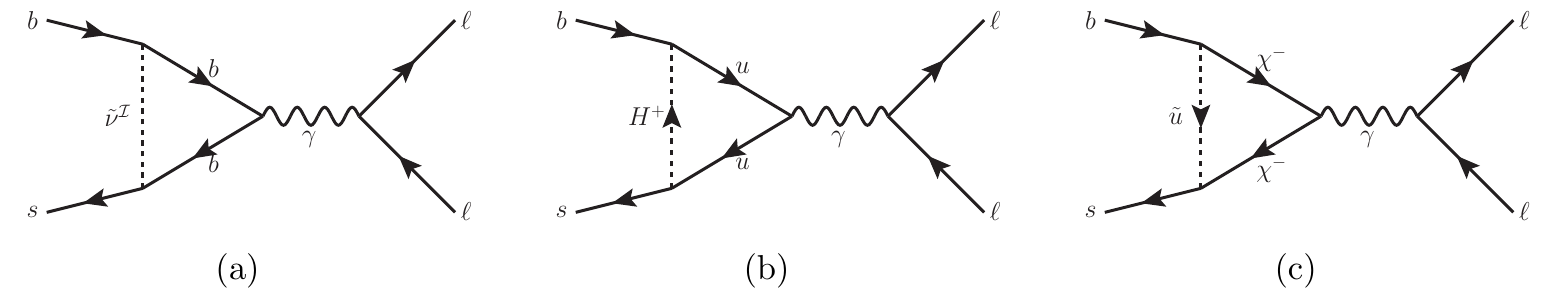}
	\caption{Photon penguin diagrams for $b \rightarrow s \ell^+\ell^-$ process in our parameter scheme. Here each example of the $\lambda'$ diagrams (left) and the non-$\lambda'$ diagrams with $W/H^\pm$ (middle) and charginos (right) engaged are shown respectively.}
	\label{fig:pengdiagram}
\end{figure}

Next we show the penguin contributions. Indeed, the Wilson coefficients of $Z$-boson penguin diagrams are negligible. While the contributions of photon penguin diagrams can be non-ignorable, 
where the $\lambda'$ diagrams (figure~\ref{fig:pengdiagram}a) give
\begin{align}\label{eq:gammapeng}
C^{\gamma(1)}_{\rm U} = -\frac{\sqrt{2}\lambda'^{\cal I}_{v33}\lambda'^{\cal I \ast}_{v23}}{36 G_F \eta_t m_{\tilde{\nu}^{\cal I}_v}^2} \biggl(\frac{4}{3} + \log\frac{m_b^2}{m_{\tilde{\nu}^{\cal I}_v}^2}  \biggr),
\end{align}
for the case of $k=3$ and the non-$\lambda'$ contributions (figure~\ref{fig:pengdiagram}b and c) namely as $C^{\gamma(2)}_{\rm U}$ are also calculated by us for completeness.

\section{The $(g-2)_{\ell}$ and other constraints}\label{sec:amu&constraints}
In this section, we introduce the NP contributions to $(g-2)_{\ell}$ and other related processes.

\subsection{The muon (electron) anomalous magnetic moment}
\label{sec:g-2}
The amplitude of the $\ell \rightarrow \ell \gamma$ ($\ell=e,\mu$) transition is given by
\begin{align}
i{\cal M}=ie \bar{\ell} \left(\gamma^\eta + a_\ell \frac{i\sigma^{\eta\beta}q_\beta}{2m_\ell} \right) \ell A_\eta 
\end{align}
in the zero limit of photon moment $q$. The second term in the bracket gives the loop corrections and $a_\ell$ is called the anomalous magnetic moment for the related lepton.

The SM-like diagrams that only involve SM particles give the same  contributions to $a_{\ell}$ as the SM. Hence the SUSY part can contribute to the observed anomaly $\Delta a_{\ell}$~\cite{Cao:2021lmj}. 
The one-loop chargino and neutralino contributions in RPV-MSSMIS are given here, with referring to Refs.~\cite{Moroi:1995yh,Martin:2001st,Altin:2017sxx,Cao:2021lmj},
\begin{align}
&\delta a_{\ell}^{\chi^\pm}  =  \frac{m_\ell}{16\pi^2}
\left[  \frac{m_\ell}{ 6 m^2_{\tilde\nu_v}}
   \left(|c^{\ell L}_{mv}|^2+|c^{\ell R}_{mv}|^2\right) F^C_1(m^2_{\chi^\pm_m}/m^2_{\tilde{\nu}_v})
   +\frac{m_{\chi^\pm_m}}{m^2_{\tilde\nu_v}}
         {\rm Re}( c^{\ell L}_{mv} c^{\ell R}_{mv}) F^C_2(m^2_{\chi^\pm_m}/m^2_{\tilde{\nu}_v})
   \right],
   \notag \\
&\delta a_\ell^{\chi^0}  = \frac{m_\ell}{16\pi^2}
\left[ -\frac{m_\ell}{ 6 m^2_{\tilde{l}_i}}
  \left( |n^{\ell L}_{ni}|^2+ |n^{\ell R}_{ni}|^2 \right) F^N_1(m^2_{\chi^0_n}/m^2_{\tilde{l}_i})
+ \frac{m_{\chi^0_n}}{ m^2_{\tilde{l}_i}}
    {\rm Re}(n^{\ell L}_{ni} n^{\ell R}_{ni}) F^N_2(m^2_{\chi^0_n}/m^2_{\tilde{l}_i})  
   \right],
\end{align}
where
\begin{align}\label{eq:alCoef}
&c^{\ell R}_{mv}  =  y_\ell U_{m2} \tilde{\cal V}^{\cal I }_{v\ell}, \quad
c^{\ell L}_{mv}  = - g_2 V_{m1} \tilde{\cal V}^{\cal I }_{v\ell}+V_{m2} Y^{\cal I}_{\ell v}; \notag\\
&n^{\ell R}_{ni}  = \sqrt{2} g_1 N_{n1} \delta_{i(\ell+3)} + y_{\ell} N_{n3} \delta_{i\ell},
\quad
n^{\ell L}_{ni}  = \frac{1}{\sqrt{2}} \left (g_2 N_{n2} + g_1 N_{n1} \right) \delta_{i\ell}
-y_\ell N_{n3} \delta_{i(\ell+3)},
\end{align}
and functions
\begin{align}
&F^C_1(x) =\frac{1}{(1-x)^4}\left( 2+ 3x - 6x^2 + x^3 +6x\log x\right), \notag\\
&F^C_2(x) =-\frac{1}{(1-x)^3}\left( 3-4x+x^2+2\log x\right), \notag\\
&F^N_1(x) = \frac{1}{(1-x)^4}\left( 1-6x+3x^2+2x^3-6x^2\log x\right),  \notag\\
&F^N_2(x)  = \frac{1}{(1-x)^3}\left( 1-x^2+2x\log x\right).
\end{align}
The flavour mixing of RH sleptons is not considered here, and neither is the flavour mixing of LH sleptons. Since the contributions from $\lambda'$ diagrams are negligible for the heavy $\tilde{b}_R$ and $\tilde{u}_L$, they are not shown in Eq.~\eqref{eq:alCoef}.
The difference from MSSM is the form factor $c^{\ell L}_{mv}$ in Eq.~\eqref{eq:alCoef}, where the extra $V_{m2} Y^{\cal I}_{\ell v}$ term can make an enhancement. Because the measured $\Delta a_{e}$ has different features compared with $\Delta a_{\mu}$, we consider the scheme of $|\delta a_{\mu}^{\chi^\pm}| \gg |\delta a_{e}^{\chi^\pm}| \approx 0$ and $|\delta a_{e}^{\chi^0}| \gg |\delta a_{\mu}^{\chi^0}| \approx 0$~\cite{Badziak:2019gaf}. And thus, the muon generation associated with RH  sleptons is set sufficiently heavy as well as heavy ${\tilde{L}'_1}$ which is already assumed in section~\ref{sec:bsll}. Afterwards we expect $1.92(1.33)\leqslant|\delta a_{\mu}^{\chi^\pm}+\delta a_{\mu}^{\chi^0}|\times 10^9 \leqslant 3.10(3.69)$ to be in accord with $a_\mu$ data at $1(2)\sigma$ respectively. 

\subsection{Tree-level processes}\label{sec:treeconstraints}
In the following we investigate related transition bounds which exchange $\tilde{d}_{Rk}$ at the tree level. On account of the assumption of heavy $m_{\tilde{d}_{Rk}}\sim 10$ TeV, the neutral current processes $B \to K^{(\ast)} \nu \bar\nu$, $B \to \pi \nu \bar\nu$, $D^0 \to \mu^+ \mu^-$ as well as the charged current processes $B \to \tau \nu$, $D_s \to \tau \nu$ and $\tau \to K \nu$ provide no effective constraints. Besides, there are some tree-level processes which may make some slight bounds to mention here.

The SM prediction ${\cal B}(K^+ \to \pi^+ \nu \bar\nu)_{\rm SM}=(9.24\pm0.83)\times 10^{-11}$~\cite{Aebischer:2018iyb} combined with the experimental measurement ${\cal B}(K^+ \to \pi^+ \nu \bar\nu)_{\rm exp}=(1.7\pm1.1)\times 10^{-10}$~\cite{ParticleDataGroup:2020ssz} induces the  constraint. The effective Lagrangian for $K \to \pi \nu \bar\nu$ decay can be described by
\begin{align}
{\cal L}_{\rm eff} = (C^{\rm SM} \delta_{ii'} + C^{\rm NP}_{ii'}) ({\bar d} \gamma_\mu P_L s)({\bar \nu}_i \gamma^\mu P_L \nu_{i'}) + {\rm h.c.},
\end{align}
where the SM contributes $C^{\rm SM}=- \frac{\sqrt{2} G_F e^2 K_{ts}K^\ast_{td}}{4\pi^2 \sin^2\theta_W} X(x_t)$ and $X(x_t) = \frac{x_t(x_t+2)}{8(x_t-1)}+\frac{3 x_t (x_t-2)}{8(x_t-1)^2}\log(x_t)$ with $x_t = m^2_t/m^2_W$~\cite{Buras:2014fpa}. While the NP contribution is
\begin{align}
C^{\rm NP}_{ii'} =\frac{\lambda'^{\cal N}_{i'2k}\lambda'^{\cal N\ast}_{i1k}}{2 m^2_{\tilde{d}_{Rk}}}.
\end{align}
Then the bound is obtained as $|\lambda'^{\cal N}_{i'2k}\lambda'^{\cal N\ast}_{i1k}|<0.074$ when $m_{\tilde{d}_{Rk}}$ around $10$ TeV~\cite{Hu:2020yvs}, and hence, we can set $|\lambda'_{i1k}|\lesssim 10^{-2}$ to avoid this bound. 

As to the processes of $\tau$ decaying to a $\mu$ and a vector meson, $\tau\rightarrow \mu \rho^0$ and $\tau\rightarrow \mu \phi$, the branching fraction is given by~\cite{Kim:1997rr},
\begin{align}
{\cal B}(\tau \to \mu V) = 
\frac{\tau_\tau f^2_V m^3_\tau}{128\pi}|A_V|^2 \left(1-\frac{m^2_V}{m^2_\tau}\right)
\left(1+ \frac{m^2_V}{m^2_\tau} - 2\frac{m^4_V}{m^4_\tau}\right),
\end{align} 
where $V$ stands for $\rho^0$ and $\phi$. The mean lifetime of tauon $\tau_\tau=290.3\pm 0.5$ fs~\cite{ParticleDataGroup:2020ssz} and the decay constants $f_V$ have the value of $f_{\rho^0}=153$ MeV and $f_\phi=237$ MeV respectively~\cite{Earl:2018snx}. And $A_V$ is given by~\cite{Kim:1997rr}
\begin{align}
&A_{\rho^0} = \frac{\tilde{\lambda}'_{3j1} \tilde{\lambda}'^{\ast}_{2j1}}{2m^2_{\tilde{u}_{Lj}}} - \frac{\tilde{\lambda}'_{31k}\tilde{\lambda}'^{\ast}_{21k}}{2m^2_{\tilde{d}_{Rk}}}, \notag\\
&A_{\phi} = \frac{\tilde{\lambda}'_{3j2} \tilde{\lambda}'^{\ast}_{2j2}}{2m_{\tilde{u}_{Lj}}^2}.
\end{align}
The most recent experimental upper limits on the branch fractions of the two processes at $90\%$ confidence level (CL) are ${\cal B}(\tau \rightarrow \mu \rho^0) < 1.2 \times 10^{-8}$ and ${\cal B}(\tau \rightarrow \mu \phi) < 8.4 \times 10^{-8}$~\cite{ParticleDataGroup:2020ssz}.  We can obtain the bounds~\cite{Earl:2018snx}
\begin{align}\label{eq:cons_taumuV}
&|\tilde{\lambda}'_{3j1} \tilde{\lambda}'^{\ast}_{2j1}- \tilde{\lambda}'_{31k}\tilde{\lambda}'^{\ast}_{21k} |<1.9, \notag\\
&|\tilde{\lambda}'_{3j2} \tilde{\lambda}'^{\ast}_{2j2} |<3.6,
\end{align}  
respectively when both $m_{\tilde{d}_{Rk}}$ and $m_{\tilde{u}_{Lj}}$ around $10$ TeV. Under the negligible $|\lambda'_{i1k}|$ assumption, the bounds of Eq.~\eqref{eq:cons_taumuV} turn into
$|\tilde{\lambda}'_{321} \tilde{\lambda}'^{\ast}_{221}+\tilde{\lambda}'_{331} \tilde{\lambda}'^{\ast}_{231} |<1.9$ or 
$|\tilde{\lambda}'_{322} \tilde{\lambda}'^{\ast}_{222}+\tilde{\lambda}'_{332} \tilde{\lambda}'^{\ast}_{232} |<3.6$
when the single value $k$ is restricted to $1$ or $2$ for the nonzero $\lambda'_{ijk}$, respectively.
While there exists no effective bound when $k$ is restricted to $3$. We can see that if $k$ is restricted to $1(2)$ for the nonzero $\lambda'_{ijk}$, $|\lambda'_{ij1(2)}|$ should be below around $1(1.3)$, respectively, in the case of no cancelling out.

\subsection{$B \rightarrow X_s \gamma$}
At one-loop level, the photon penguin diagrams in figure~\ref{fig:pengdiagram} also contribute to the electromagnetic dipole operator ${\cal O}_7=\frac{m_b}{e} (\bar{s} \sigma^{\mu\nu} P_{R} b) F_{\mu\nu}$ constrained by the $B\rightarrow X_s \gamma$ decay. The NP Wilson coefficient $C^{\rm NP}_7$ includes charged Higgs contributions and the effects from the chargino with $\tilde{u}_{Rj}$ as well as the RPV contributions engaging sneutrinos.

The RPV contribution is given by
\begin{align}
C^{\rm RPV}_7=\frac{\sqrt{2}\lambda'^{\cal I}_{v3k}\lambda'^{\cal I \ast}_{v2k}}{144  G_F \eta_t m_{\tilde{\nu}^{\cal I}_v}^2}.
\end{align}
Compared with $C^{\gamma(1)}_{\rm U}$ in Eq.~\eqref{eq:gammapeng}, $C^{\rm RPV}_7$ contains the common part $\lambda'^{\cal I}_{v3k}\lambda'^{\cal I \ast}_{v2k}/m_{\tilde{\nu}^{\cal I}_v}^2$ while no logarithmic term.

The contribution of charged Higgs is given by
\begin{align}
C^{H^\pm}_7=\frac{1}{3\tan^2\beta}F^{(1)}_7(y_t)+F^{(2)}_7(y_t),
\end{align}
where the form factors are $F^{(1)}_7(y_t)=\frac{y_t(7-5y_t-8y_t^2)}{24(y_t-1)^3}+\frac{y_t^2(3y_t-2)}{4(y_t-1)^4}\log y_t$ and $F^{(2)}_7(y_t)=\frac{y_t(3-5y_t)}{12(y_t-1)^2}+\frac{y_t(3y_t-2)}{6(y_t-1)^3}\log y_t$ with $y_t=m^2_t/m^2_{H^\pm}$. 
One can see the $F^{(2)}_7(y_t)$ part is not suppressed when $\tan\beta$ is large and this is unlike the $H^\pm$ contributions to $C_{9(10)}$ which are entirely suppressed by large $\tan\beta$. 
The formulas of $C^{\chi^\pm}_7$ engaging the chargino together with $\tilde{u}_{Rj}$ and the QCD corrections can be seen in Ref~\cite{deCarlos:1996yh}.

The recent measurements of branching ratio ${\cal B}(B\rightarrow X_s \gamma)_{\rm exp}\times 10^4=3.43\pm 0.21\pm 0.07$~\cite{Amhis:2019ckw} is consist with the SM prediction ${\cal B}(B\rightarrow X_s \gamma)_{\rm SM}\times 10^4=3.36\pm 0.23$~\cite{Misiak:2015xwa} and induce the bound to $|C^{\rm NP}_7|<0.025$~\cite{Altmannshofer:2020axr}.

\subsection{$B_s-\bar{B}_s$ mixing}
Another process we should consider is $B_s-\bar{B}_s$ mixing, mastered by the Lagrangian
\begin{align}
{\cal L}_{\rm eff}=(C_{B_s}^{\text{SM}}+C_{B_s}^{\text{NP}})
(\bar{s} \gamma_{\mu} P_L b)(\bar{s} \gamma^{\mu} P_L b)+{\rm h.c.},
\end{align}
where the non-negligible NP contribution is given by
\begin{align}
C_{B_s}^{\text{NP}} = &-\frac{i}{8} \biggl(
\lambda'^{\cal I}_{v3k}\lambda'^{\cal I \ast}_{v2k}\lambda'^{\cal I}_{v'3k}\lambda'^{\cal I \ast}_{v'2k}D_2[m_{\tilde{\nu}^{\cal I}_v},m_{\tilde{\nu}^{\cal I}_{v'}},m_{d_k},m_{d_k}] \notag \\
&+ y^2_{u_{i}} y^2_{u_{j}} 
(K_{i3}K^{\ast}_{i2})(K_{j3}K^{\ast}_{j2})|V_{m2}V_{n2}|^2 
D_2[m_{\chi^\pm_m},m_{\chi^\pm_n},m_{\tilde{u}_{Ri}},m_{\tilde{u}_{Rj}}]
\biggr)
\end{align}
including the $\lambda'$ diagram with double sneutrinos and the non-$\lambda'$ diagram with double RH su-quarks, and the SM contribution 
$C_{B_s}^{\rm SM} =-\frac{1}{4 \pi^2} G_F^2 m_W^2 \eta_t^2 S(x_t)$ with the function $S(x_t)=\frac{x_t(4-11x_t+x_t^2)}{4(x_t-1)^2}+\frac{3x_t^3\log(x_t)}{2(x_t-1)^3}$. With the measurement of $\Delta M_s^{\rm exp}=(17.757 \pm 0.021)$ ${\rm ps}^{-1}$~\cite{Amhis:2019ckw}\footnote{The newly  updated experimental result of $\Delta M_s$ by LHCb has been reported~\cite{Aaij:2021jky}. The combined result with previous LHCb measurements gives $\Delta M_s^{\rm LHCb}=(17.7656 \pm 0.0057)$ ${\rm ps}^{-1}$ with the improved precision. Using this new combined result will not change the Eq.~\eqref{eq:DMsbound}.}, the recent SM prediction $\Delta M_s^{\rm SM}=(18.4^{+0.7}_{-1.2})~{\rm ps}^{-1}=(1.04^{+0.04}_{-0.07})\Delta M_s^{\rm exp}$~\cite{DiLuzio:2019jyq}   
leads to the bound of 
\begin{align}\label{eq:DMsbound}
0.90 < |1+ C_{B_s}^{\rm NP}/C_{B_s}^{\rm SM}| < 1.11,
\end{align}
at $2\sigma$ level.

\subsection{Lepton flavour violating decays}\label{sec:LFVdecay}
We discuss the lepton flavour violating decays including $\tau \rightarrow \ell \gamma$, $\mu \rightarrow e \gamma$, $\tau \rightarrow \ell \ell \ell$, $\mu \rightarrow e e e$ and $\tau \rightarrow \ell'  \ell \ell$. 
Firstly, the $\lambda'$-diagram contributions 
can be eliminated when $\tilde{b}_R$ is sufficiently heavy~\cite{Hu:2020yvs}. As to the non-$\lambda'$ diagrams, all the neutralino-slepton diagrams contain flavour mixings of charged sleptons and all the chargino-sneutrino diagrams contain flavour mixings of sneutrinos (see Ref.~\cite{Abada:2014kba} for concrete formulas). So the effects of these two kinds of diagrams vanish when there are no flavour mixing in the two  mass matrices. For contributions of $W/H^\pm$-neutrino diagrams, they are always connected to these terms which are ${\cal V}^{T\ast}_{(\alpha+3)v}{\cal V}^{T}_{(\beta+3)v}$, ${\cal V}^{T\ast}_{(\alpha+3)v}{\cal V}^{T}_{\beta v}$ and ${\cal V}^{T\ast}_{\alpha v}{\cal V}^{T}_{\beta v}$ and their conjugate terms, where $\alpha,\beta=e,\mu,\tau$ and $\alpha\neq\beta$~\cite{Abada:2014kba}. In section~\ref{sec:inputs}, we will show all these terms contribute no effects under the particular structure of neutrino mass matrix. The same analyses can also be applied to the non-$\lambda'$ diagrams in $B^0_s \rightarrow \tau^\pm \mu^\mp$ and $B^+\rightarrow K^+\tau^\pm \mu^\mp$ . For the $\lambda'$ diagrams of these two processes, we refer to the detailed discussions in Ref.~\cite{Hu:2019ahp}, and no obvious constraints are found.

\subsection{Anomalous $t \rightarrow c V(h)$ decays}\label{sec:tcX}
The SM predicts the branching ratios of $t\rightarrow c V$ decays ($V$ stands for the vector bosons including $Z$, $\gamma$ and the gluon $g$) and $t\rightarrow c h$ decay ($h$ stands for SM-like Higgs) below the scale of $10^{-15} - 10^{-12}$~\cite{TopQuarkWorkingGroup:2013hxj} due to the Glashow–Iliopoulos–Maiani suppression. This scale is beyond the detection capabilities at the collider in the near future. 
The most recent experimental $95\%$ CL upper limits on the branching ratios of these top quark decays at the Large Hadron Collider (LHC) show that ${\cal B}(t\rightarrow cZ)<2.4\times10^{-4}$~\cite{ATLAS:2018zsq}, ${\cal B}(t\rightarrow c\gamma)<1.8\times10^{-4}$~\cite{ATLAS:2019mke}, ${\cal B}(t\rightarrow cg)<4.1\times10^{-4}$~\cite{CMS:2016uzc} and ${\cal B}(t\rightarrow ch)<1.1\times10^{-3}$~\cite{ATLAS:2018jqi}. 
Compared with the effects from pure MSSM, the one-loop RPV diagrams can make more contributions~\cite{Yang:1997dk,Eilam:2001dh}, and hence we will investigate these effects in our model.

For the $t\rightarrow cV$ decays, the effective $tcV$ vertices are expressed as,
\begin{align}
&V^{\mu}(tcZ)=ie\left(\gamma^{\mu}P_L A^Z
              +ik_{\nu}\sigma^{\mu\nu}P_R B^Z \right), \notag\\
&V^{\mu}(tc\gamma)=ie\left(ik_{\nu}\sigma^{\mu\nu}P_R B^{\gamma}\right),\notag\\
&V^{\mu}(tcg)=ig_s t^a\left(ik_{\nu}\sigma^{\mu\nu}P_R B^g \right),
\end{align}
where $k_{\nu}$ is the momentum of the vector boson. The form factors $A^Z$ and $B^V$ are given by~\cite{Yang:1997dk}
\begin{align}
A^Z=&\frac{\tilde{\lambda}'^{\ast}_{i2k}\tilde{\lambda}'_{i3k}}{16\pi^2}
\Bigl\{
f^Z_1 B_1(m_t,m_{d_k},m_{\tilde{l}_{Li}})-f^Z_2[2c_{24}-\frac{1}{2}+m^2_Z(c_{12}+c_{23})](-p_t,p_c,m_{d_k},m_{\tilde{l}_{Li}},m_{d_k}) \notag\\
&-f^Z_3[2c_{24}+m^2_t(c_{11}-c_{12}+c_{21}-c_{23})](-p_t,k,m_{d_k},m_{\tilde{l}_{Li}},m_{\tilde{l}_{Li}}) \Bigr\}, \notag\\
B^V=&\frac{\tilde{\lambda}'^{\ast}_{i2k}\tilde{\lambda}'_{i3k}}{16\pi^2}
\Bigl\{
f^V_2 m_t[c_{11}-c_{12}+c_{21}-c_{23}](-p_t,p_c,m_{d_k},m_{\tilde{l}_{Li}},m_{d_k}) \notag\\
&-f^V_3 m_t[c_{11}-c_{12}+c_{21}-c_{23}](-p_t,k,m_{d_k},m_{\tilde{l}_{Li}},m_{\tilde{l}_{Li}}) \Bigr\},
\end{align}
where $p_t$ and $p_c$ are the momentums of top and charm quarks and functions $B_1$ and $c_{ij}$ are the Passarino–Veltman integrals totally referring to Ref.~\cite{Passarino:1978jh}. The constants $f^Z_1=\frac{3-4\sin^2\theta_W}{6\sin\theta_W\cos\theta_W}$, $f^Z_2=-\frac{\sin\theta_W}{3\cos\theta_W}$ and $f^Z_3=\frac{2\sin^2\theta_W-1}{\sin\theta_W\cos\theta_W}$ are in the form factors $A^Z$ and $B^Z$. And in $B^\gamma$, the relevant constants are $f^\gamma_2=1/3$ and $f^\gamma_3=-1$. While $f^g_2=-1$ and $f^g_3=0$ in $B^g$. Then the decay widths of $t\rightarrow c V$ are given
\begin{align}
&\Gamma(t\rightarrow cZ)_{\rm NP}=\frac{e^2(m^2_t-m^2_Z)^2}{32\pi m^3_t}\left[(2+\frac{m^2_t}{m^2_Z})|A^Z|^2-6m_t{\rm Re}(A^ZB^{Z\ast})+(2m^2_t+m^2_Z)|B^Z|^2\right], \notag\\
&\Gamma(t\rightarrow c\gamma)_{\rm NP}=\frac{e^2m^3_t}{16\pi}|B^\gamma|^2, \notag\\
&\Gamma(t\rightarrow cg)_{\rm NP}=\frac{g^2_s m^3_t}{12\pi}|B^g|^2.
\end{align}

As for the $t\rightarrow ch$ decay, the effective $tch$ vertex is expressed as, 
\begin{align}
V(tch)=ie \left(P_L A^h_L+P_R A^h_R \right).
\end{align}  
After omitting masses of charm quarks and all down-type quarks, one can obtain~\cite{Eilam:2001dh}
\begin{align}
&A^h_R=\frac{\tilde{\lambda}'^{\ast}_{i2k}\tilde{\lambda}'_{i3k}}{16\pi^2}
{\cal Y}_{\tilde{l}_{Li}} m_t(c_{11}-c_{12})(-p_t,k,0,m_{\tilde{l}_{Li}},m_{\tilde{l}_{Li}}), \notag\\
&A^h_L=0,
\end{align}
where factor ${\cal Y}_{\tilde{l}_{Li}}\approx \frac{m_Z}{\sin\theta_W\cos\theta_W}(\frac{1}{2}-\sin^2\theta_W)\cos 2\beta$ when the masses of leptons are omitted and the mass of $CP$-odd Higgs is sufficiently heavy. Then the decay width of $t\rightarrow ch$ is
\begin{align}
\Gamma(t\rightarrow ch)_{\rm NP}=\frac{e^2(m^2_t-m^2_h)^2}{32\pi m^3_t}|A^h_R|^2.
\end{align}

The NP contributes to related branching ratios are given by ${\cal B}(t\rightarrow cV(h))_{\rm NP}=\Gamma(t\rightarrow cV(h))_{\rm NP}/\Gamma(t\rightarrow b W)_{\rm SM}$ where the dominant decay $t\rightarrow b W$ has the SM prediction $\Gamma(t\rightarrow b W)_{\rm SM}=1.42$~GeV~\cite{Beneke:2000hk}.

\subsection{LHC direct searches}\label{sec:LHCbound}
The LHC direct searches have led to stringent limits on the masses of sbottoms and stops~\cite{Aaboud:2017opj,CMS:2018qxv,ATLAS:2019gqq,ATLAS:2020xyo,ATLAS:2021fbt,CMS:2021beq,CMS:2021eha} and the recent searches~\cite{ATLAS:2021fbt,CMS:2021beq,CMS:2021eha} have excluded the heavy  stops more than $1.35$ TeV. In view of that $\tilde{d}_{Rk}$ and $\tilde{u}_{Li}$ have already been set adequately heavy, we further set $m_{\tilde{u}_{Ri}}>1.4$ TeV. 

For the constraints on LH sleptons as mentioned in section~\ref{sec:model}, when considering non-zero $\lambda$ couplings, the lower bounds of $m_{\tilde{\nu}_L}$ and $m_{\tilde{l}}$ reach TeV scale~\cite{ATLAS:2018rns,ATLAS:2018mrn,ATLAS:2021yyr}. Because only non-zero $\lambda'$ couplings are restricted in this work, LH sleptons decaying to pure leptons directly is secondary and processes without $\lambda$ interactions should been taken into consideration mainly. We consider the searches which focus on LH sleptons decaying into leptons and the lightest neutralino $\chi^0_1$~\cite{CMS:2018eqb,ATLAS:2019lff,CMS:2020bfa}, and the recent ATLAS results~\cite{ATLAS:2019lff} show that the LH sleptons which are heavier than $\chi^0_1$ can avoid the exclusion for $m_{\chi^0_1}\gtrsim 300$ GeV. Besides, the compressed scenario that the lightest chargino mass $m_{\chi^\pm_1}$ is slightly heavier than $m_{\chi^0_1}$~\cite{ATLAS:2019lng}, is adopted. Thus we will let inputs inducing $m_{\chi^\pm_1}\gtrsim m_{\chi^0_1}\gtrsim 300$ GeV and $m_{\tilde{l}_L}>300$ GeV.   

\section{Numerical results and discussions}\label{sec:num}
In this section, we investigate $b\rightarrow s\ell^+\ell^-$ anomalies numerically as well as the $a_\mu$ anomaly and the related constraints.

\subsection{Choice of input parameters}\label{sec:inputs}
\begin{table}[t]
	\setlength\tabcolsep{5pt}
	\renewcommand\arraystretch{1.3}
	\begin{center}
		\vspace{0.18cm}
		\begin{tabular}{|c|c|c|c|c|c|}
			\hline
			\multicolumn{6}{|l|}{QCD and EW parameters~\cite{ParticleDataGroup:2020ssz}}\\
			\hline
			$G_F[10^{-5}~\text{GeV}^{-2}]$ & $\alpha_e(m_Z)$ & $\alpha_s(m_Z)$  &  $m_W[\text{GeV}]$ & \multicolumn{2}{c|}{$\sin^2\theta_W$}\\
			\hline
			$1.1663787$ & ${1}/{128}$ & $0.1179(10)$ &  $80.379$ & \multicolumn{2}{c|}{$0.2312$}\\
			\hline
			\multicolumn{6}{|l|}{Quark and lepton masses [GeV]~\cite{ParticleDataGroup:2020ssz}}\\
			\hline
			$\overline{m}_b(\overline{m}_b)$ & $\overline{m}_c(\overline{m}_c)$ & $m_t$ & $m_e$ & {$m_\mu$} & $m_\tau$ \\
			\hline
			$4.18^{+0.03}_{-0.02}$ & $1.27(2)$ & $172.76(30)$ & $0.511\times 10^{-3}$ & {$0.1057$} & $1.777$  \\
			\hline
			\multicolumn{6}{|l|}{CKM Wolfenstein parameters~\cite{CKMfitterwebsite}} \\
			\hline
			$\lambda_{\text{CKM}}$ & $A$ & $\bar{\rho}$ & \multicolumn{3}{c|}{$ \bar{\eta}$} \\
			\hline
			$0.22484^{+0.00025}_{-0.00006}$ & $0.8235^{+0.0056}_{-0.0145}$ & $0.1569^{+0.0102}_{-0.0061}$ & \multicolumn{3}{c|}{$0.3499^{+0.0079}_{-0.0065}$} \\
			\hline 
			\multicolumn{6}{|l|}{Lepton oscillation parameters (NO)~\cite{Esteban:2020cvm}} \\
			\hline
			\multicolumn{1}{|c}{$\sin^2\theta_{12}$} & \multicolumn{2}{|c|}{$\sin^2\theta_{23}$} & \multicolumn{3}{c|}{$\sin^2\theta_{13}$}  \\
			\hline
			\multicolumn{1}{|c}{$0.304(12)$} & \multicolumn{2}{|c|}{$0.573^{+0.016}_{-0.020}$} & \multicolumn{3}{c|}{$0.02219^{+0.00062}_{-0.00063}$} \\
			\hline
			\multicolumn{1}{|c}{$\delta_{\rm CP}[^{\circ}]$} & \multicolumn{2}{|c|}{$\Delta m^2_{21}[10^{-5}~\text{eV}^{2}]$} & \multicolumn{3}{c|}{$\Delta m^2_{31}[10^{-3}~\text{eV}^{2}]$} \\
			\hline
			\multicolumn{1}{|c}{$197^{+27}_{-24}$} & \multicolumn{2}{|c|}{$7.42^{+0.21}_{-0.20}$} & \multicolumn{3}{c|}{$2.517^{+0.026}_{-0.028}$} \\
			\hline
		\end{tabular}\caption{Summary of parts of input parameters used throughout this paper.}\label{tab:input1}
	\end{center}
\end{table}

\begin{table}[t]
\centering
\setlength\tabcolsep{8pt}
\renewcommand{\arraystretch}{1.3}
\begin{tabular}{|cc||cc|}
        \hline
		Parameters & Sets & Parameters  & Sets \\
		\hline
  $\tan\beta$  & $15$      & $Y_\nu$   & $\text{diag}(0.7,0.8,0.5)$~\cite{DeRomeri:2018pgm}\\
		$M_1$  & $320$ GeV & $M_R$     & $\text{diag}(1,1,1)$ TeV \\
		$M_2$  & $350$ GeV & $B_{M_R}$ & $\text{diag}(0.5,0.5,0.5)~\text{TeV}^2$    \\
		$\mu$  & $450$ GeV & $A_\nu$   & $0$;~$\text{diag}(0,-1.5,0)$ TeV \\
		\hline
$m_{\tilde{u}_{Ri}}$ & $1.5$ TeV & $m_{\tilde{L}'_1}$ & $5$ TeV \\
$m_{\tilde{\mu}_R}$ &  $5$ TeV   & $m_{\tilde{R}}$ & $\text{diag}(5,0,0)$ TeV \\
    \hline
	\end{tabular}
	\caption{The sets of fixed model parameters, defined at $\mu_{\rm NP}$ scale. The two sets of $A_\nu$ are for scenario A and B respectively.}
	\label{tab:input2}
\end{table}

First parts of input parameters used throughout the paper are collected in table~\ref{tab:input1}, which includes the lepton oscillation data~\cite{Esteban:2020cvm} under the normal ordering (NO) assumption of LH neutrino masses. In addition, we further keep $\delta_{\rm CP}=\pi$ to omit the $CP$ violation in $U^{\rm PMNS}$. The lightest neutrino mass is set zero to have the masses of three-flavour light neutrinos as $\{0, 0.008, 0.05\}$ eV with $m^{\text{diag}}_{\nu}\approx\text{diag}(0, \sqrt{\Delta m^2_{21}}, \sqrt{\Delta m^2_{31}})$~\cite{Alvarado:2020lcz}. Then we collect the fixed values of relevant model parameters in table~\ref{tab:input2}. The sets give $m_{\chi^\pm_1}=325$ GeV and $m_{\chi^0_1}=307$ GeV which are in accord with the constraints discussed in section~\ref{sec:LHCbound}. Besides, we set the diagonal parameters for $Y_\nu$, $M_R$,  $m_{\tilde{L}'}$, $m_{\tilde{R}}$, $B_{M_R}$ and $A_{\nu}$ in Eq.~\eqref{eq:mSnuNum} so that the sneutrino mixing matrices $\tilde{\cal V}^{\cal I(R)}$ only have chiral mixings without flavour mixings and let them have proper values to agree with the discussions in sections~\ref{sec:bsll} and \ref{sec:amu&constraints}
\footnote{
Under the premise of no flavour mixing in $\tilde{\cal V}^{\cal I(R)}$, we mention that sufficiently heavy $m_{\tilde{L}'_1}$ and $m_{\tilde{R}_1}$ can eliminate the box contributions to $b\rightarrow se^+e^-$ except for $C^{W/H^\pm(1)}_{9,e}$ in section~\ref{sec:bsll}; $m_{\tilde{\mu}_R}$ and $m_{\tilde{L}'_1}$ are set sufficiently heavy for the scheme of non-dominant $|\delta a_{\mu}^{\chi^0}|$ and $|\delta a_{e}^{\chi^\pm}|$ in section~\ref{sec:g-2}. The $\tilde{\cal V}^{\cal I(R)}$ without flavour mixings is also for satisfying the bounds from lepton flavour violating decays mentioned in section~\ref{sec:LFVdecay}.}. 
As for the remaining model parameters, $m_{\tilde{L}'_{2}}$, $m_{\tilde{L}'_{3}}$, $\lambda'_{22k}$, $\lambda'_{23k}$, $\lambda'_{32k}$ and $\lambda'_{33k}$, they can vary freely in the ranges considered. 

With related inputs in table~\ref{tab:input1} and table~\ref{tab:input2}, the $\mu_S$ term in Eq.~\eqref{eq:mnu} can be figured out with $m_D$, $M_R$, $U^{\rm PMNS}$ and the light neutrino masses $m^{\text{diag}}_{\nu}$ through Eq.~\eqref{eq:musterm}. Then we obtain the approximate numerical form of the mixing matrix ${\cal V}^T$,
\begin{align}\label{eq:exactVT}
{\cal V}^T \approx 
\left(
\begin{array}{ccccccccc}
{0.840} & {0.509} & {-0.147} & -0.085 i & 0 & 0 & 0.085 & 0 & 0 \\
{-0.231} & {0.599} & {0.755} & 0 & 0.097 i & 0 & 0 & 0.097 & 0 \\
{0.478} & {-0.608} & {0.628} & 0 & 0 & 0.061 i & 0 & 0 & -0.061 \\
0 & 0 & 0 & 0.707 i & 0 & 0 & 0.707 & 0 & 0 \\
0 & 0 & 0 & 0 & -0.707 i & 0 & 0 & 0.707 & 0 \\
0 & 0 & 0 & 0 & 0 & -0.707 i & 0 &
0 & -0.707 \\
-0.102 & -0.062 & 0.018 & -0.702 i & 0 & 0 & 0.702 & 0 & 0 \\
0.032 & -0.083 & -0.105 & 0 & 0.700 i & 0 & 0 & 0.700 & 0 \\
-0.041 & 0.053 & -0.055 & 0 & 0 & 0.704 i & 0 & 0 & -0.704 \\
\end{array}
\right),
\end{align}
corresponding to the light neutrino masses $m_{\nu_i}=\{0, 0.008, 0.05\}$ eV and the nearly degenerate heavy ones $m_{\nu_{v_h}}$ around $1$ TeV. 
One can see that there is no flavour mixing when RH sector is engaged but chiral mixing exists. With the numerical result of Eq.~\eqref{eq:exactVT}, we can eliminate the ${\cal V}^{T\ast}_{(\alpha+3)v}{\cal V}^{T}_{(\beta+3)v}$ and ${\cal V}^{T\ast}_{(\alpha+3)v}{\cal V}^{T}_{\beta v}$ terms in the contributions of $W/H^\pm$-neutrino diagrams to the lepton flavour violating decays within section~\ref{sec:LFVdecay}. ${\cal V}^{T\ast}_{\alpha v}{\cal V}^{T}_{\beta v}$ can be decomposed into the following two parts, $\sum_{v_h=4}^{9} {\cal V}^{T\ast}_{\alpha v_h}{\cal V}^{T}_{\beta v_h}$ and $\sum_{i=1}^{3} {\cal V}^{T\ast}_{\alpha i}{\cal V}^{T}_{\beta i}=-\sum_{v_h=4}^{9} {\cal V}^{T\ast}_{\alpha v_h}{\cal V}^{T}_{\beta v_h}$ related to the nearly degenerate heavy neutrinos and light neutrinos respectively~\cite{Chang:2017qgi}. It can also be found that ${\cal V}^{T\ast}_{\alpha v}{\cal V}^{T}_{\beta v}$ provides no effects from Eq.~\eqref{eq:exactVT}. In sum, the lepton flavour violating decays mentioned in section~\ref{sec:LFVdecay} contribute no effective bounds in our input sets.

In the following we discuss the feature of $C^{W/H^\pm(1)}_{9(10),\ell}$ in Eq.~\eqref{eq:WHWcs1} which includes the $e^+e^-$ channel. The three terms of Eq.~\eqref{eq:WHWcs1} contain $|{\cal V}^T_{(\ell+3)v}|^2$, ${\rm Re}({\cal V}^T_{\ell v}{\cal V}^T_{(\ell+3)v})$ and $|{\cal V}^T_{\ell v}|^2$ respectively. 
For $h$ or/and $h'=2$ firstly, this part of Eq.~\eqref{eq:WHWcs1} includes the contribution of charged Higgs and it can be ignored for the set of $\tan\beta=15$. 
In the case of $h=h'=1$, this part of Eq.~\eqref{eq:WHWcs1} describes the contribution of seesaw-extended SM. $|{\cal V}^T_{\ell v}|^2$ in the third term is dominated by $|U^{\rm PMNS}_{\ell i}|^2$, and thus, this term is nearly equal to the contribution of the original SM. Because ${\rm Re}({\cal V}^T_{\ell v}{\cal V}^T_{(\ell+3)v})$ is negligible compared with $|{\cal V}^T_{(\ell+3)v}|^2$, we focus on the first term in Eq.~\eqref{eq:WHWcs1} and the second term can be omitted safely. According to the discussion above, the pure NP contribution $\Delta C^{W/H^\pm}_{9(10),\ell}$ in Eq.~\eqref{eq:WHWcs1} is induced by only heavy neutrinos and is given by
\begin{align}\label{eq:specW}
\Delta C^{W/H^\pm}_{9,\ell}=\sum_{v_h=4}^{9} -\frac{\sqrt{2}\pi^2 i}{2G_F \eta_t e^2} 
y^2_{u_i} K_{i3}K^{\ast}_{i2} \sin^4\beta
|Y_{\nu_{\ell\ell}} {\cal V}^T_{(\ell+3)v_h}|^2 D_2[m_{\nu_{v_h}},m_{u_i},m_W,m_W].
\end{align}

Our parameter set leads that the contribution $\Delta C^{W/H^\pm}_{9,\ell}=-\Delta C^{W/H^\pm}_{10,\ell}$ in Eq.~\eqref{eq:specW} is at negative $10^{-2}$ scale for both $\mu^+\mu^-$ and $e^+e^-$ channels. This small LFU effect cannot be included in both $C_{\rm V}$ and $C_{\rm U}$ within Eq.~\eqref{eq:sce} and is ignored for the approximation. Also we obtain $C^{\gamma(2)}_{\rm U}$ around $0.01$ and $C^{\chi^\pm(1)}_{\rm V}$ around $-0.01$ in which the main contributions are from $\lambda'$ diagrams. In summary, the LFU violating coefficient and the LFU coefficient can be represented by $C_{\rm V}=C^{\chi^\pm(1)}_{\rm V}+C^{\chi^\pm(2)}_{\rm V}$ and $C_{\rm U}=C^{\gamma(1)}_{\rm U}+C^{\gamma(2)}_{\rm U}$, respectively. We find that the factor $c^{\mu L}_{mv}=-g_2 V_{m1} \tilde{\cal V}^{\cal I}_{v2}+V_{m2} Y^{\cal I}_{2v}$ in Eq.~\eqref{eq:alCoef} is also included in $C^{\chi^\pm(1)}_{\rm V}$ and $C^{\chi^\pm(2)}_{\rm V}$. Therefore, the large chiral mixing of sneutrinos will make some enhancements to both $C_{\rm V}$ and $a^{\rm NP}_\mu$ simultaneously.

\subsection{Explanations of $b\rightarrow s\ell^+\ell^-$ anomalies with $(g-2)_\mu$}
In this part we will search for the common areas of these five variables, $m_{\tilde{L}'_{2}}$, $\lambda'_{223}$, $\lambda'_{233}$, $\lambda'_{323}$ and $\lambda'_{333}$, to explain $b\rightarrow s\ell^+\ell^-$ anomalies as well as $(g-2)_\mu$ deviations considering related constraints, in two fit scenarios mentioned in section~\ref{sec:intro}. In scenario A, we fix $m_{\tilde{L}'_3}=m_{\tilde{L}'_2}-50~{\rm GeV}$ which is benefit for satisfying the constraint of $B_s-\bar{B}_s$ mixing. In scenario B, we fix $m_{\tilde{L}'_3}=m_{\tilde{L}'_2}$. For the bounds of $m_{\tilde{l}_L}$ in LHC searches mentioned in section~\ref{sec:LHCbound}, we focus on  $m_{\tilde{L}'_2}\geqslant 370$ GeV which makes the mass of lightest  charged slepton $m_{\tilde{l}_1}$ (sneutrino $m_{\tilde{\nu}_1}$) be above $318~(301)$ GeV in scenario A and $m_{\tilde{\nu}_1}$ be heavier than $100$ GeV while $m_{\tilde{l}_1}$ be above $352$ GeV in scenario B. In particular, we choose $k=3$ for a benchmark of the numerical calculation.

\subsubsection{Explanations of $(g-2)_\mu$ anomalies}
\begin{figure}[htbp]
	\centering
	\includegraphics[width=0.95\textwidth]{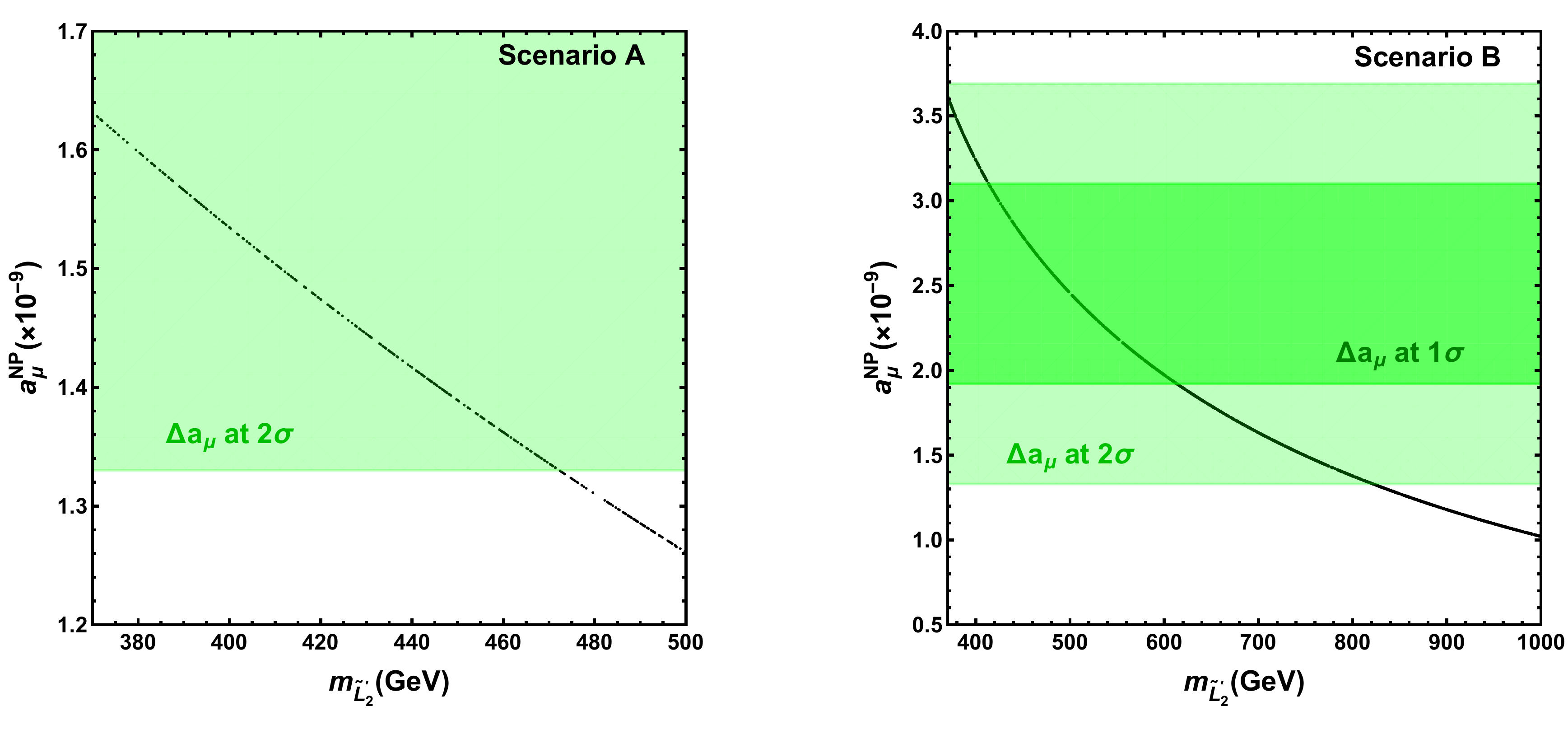}
	\caption{$a_{\mu}^{\text{NP}}$ varies with $m_{\tilde{L}^{'}_2}$ in scenario A (left) and scenario B (right). The dark (light) green areas are $1(2)\sigma$ favored to explain the $\Delta a_{\mu}$ deviation.}
	\label{fig:amu}
\end{figure}

At first, we show the bound for $m_{\tilde{L}'_2}$ with $a_{\mu}^{\text{NP}}$ in each scenario at figure~\ref{fig:amu}, in which $a_{\mu}^{\text{NP}}$ can contribute to $a_{\mu}^{\text{SM}}$ increasing and accommodate the observed $\Delta a_{\mu}$ deviation from $4.2\sigma$ to $2\sigma$ level below. In scenario B, there are also allowed spaces at $1\sigma$ level.
In scenario A, we obtain the allowed range $370 \leqslant m_{\tilde{L}'_2} \leqslant 470$~GeV at $2\sigma$ level, and in scenario B there is a larger range of $370 \leqslant m_{\tilde{L}'_2} \leqslant 820$~GeV at $2\sigma$ level as well as a range of $420 \leqslant m_{\tilde{L}'_2} \leqslant 610$~GeV at $1\sigma$ level. Besides, we have calculated that $a_e^{\text{NP}}$ can reach negative ${\cal O}(10^{-14})$ only in the case of $m_{\tilde{e}_R} \sim 100$ GeV in our parameter spaces.

\subsubsection{Results in scenario A}
\begin{figure}[htbp]
	\centering
\includegraphics[width=0.95\textwidth]{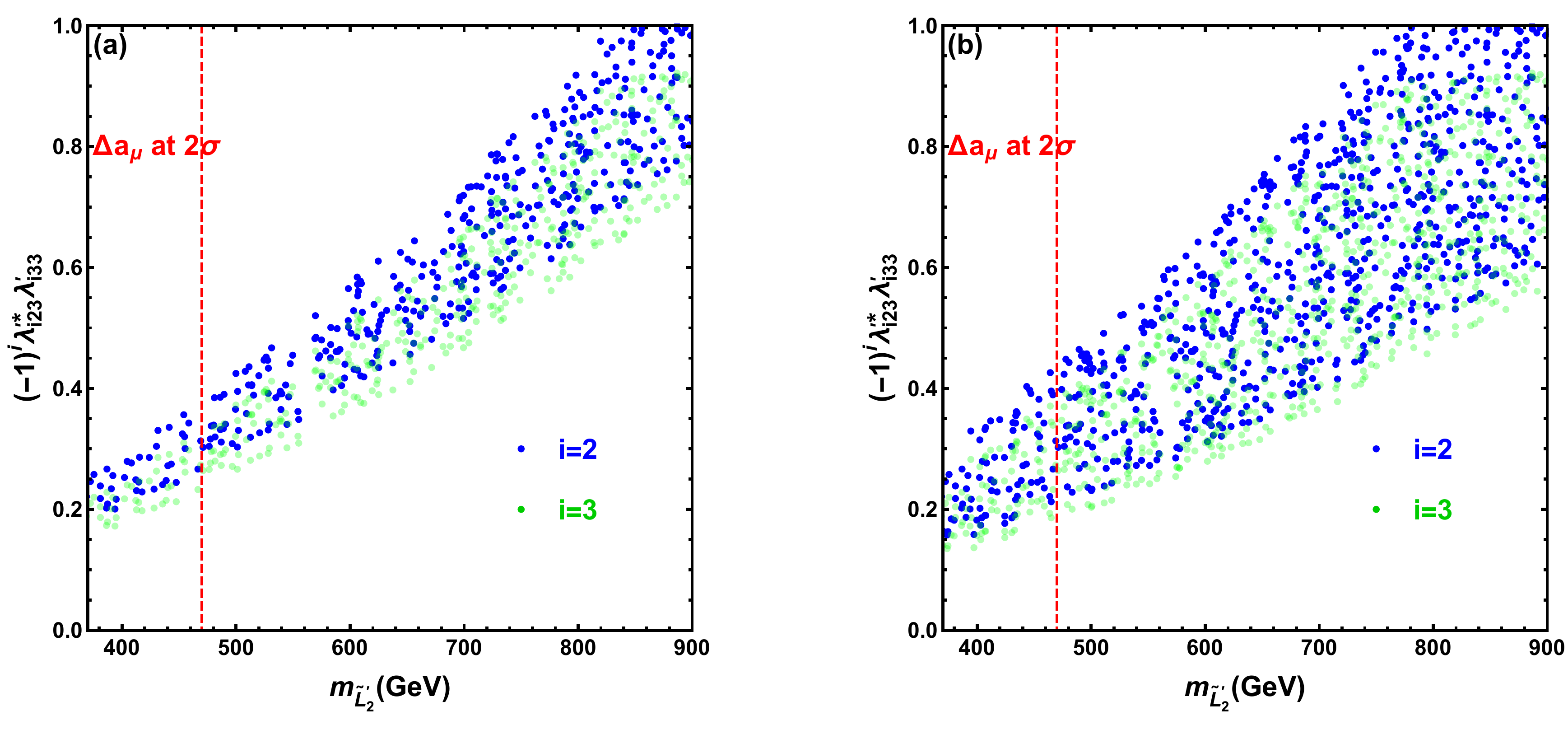}
	\caption{The common scopes of constraints with the fit level of rare $B$-meson decays as $1\sigma$ (left) and $2\sigma$ (right) in scenario A. The blue (green) points show $\lambda'^{\ast}_{223}\lambda'_{233}$ ($-\lambda'^{\ast}_{323}\lambda'_{333}$) varies with $m_{\tilde{L}'_2}$ and  then $\lambda'^{\ast}_{323}\lambda'_{333}$ ($\lambda'^{\ast}_{223}\lambda'_{233}$) is derived. 
It should be paid attention to that among $\lambda'^{\ast}_{223}\lambda'_{233}$ and $\lambda'^{\ast}_{323}\lambda'_{333}$ there is only one independent variable in this scenario so the blue and green points are relevant. 
The area on the left of the red dashed line is allowed to be accordant with $a_{\mu}$ data at $2\sigma$.}\label{fig:SceA}
\end{figure}

Next we investigate $b\rightarrow s\ell^+\ell^-$ anomalies further with the parameter regions of $m_{\tilde{L}'_2}$ we have obtained above.
In scenario A, $C_{\rm U}$ should be $0$ as the definition. 
To make $C_{\rm U}$ cancel out, $\lambda'^{\ast}_{323}\lambda'_{333}$ is figured out as the expression of $\lambda'^{\ast}_{223}\lambda'_{233}$ and vice versa, and the constraints from $B_s-\bar{B}_s$ mixing and $B\rightarrow X_s \gamma$ will be mainly suppressed. Also the large $m_{H^\pm}$ as  $2$ TeV and $m_{\tilde{u}_{Ri}}$ as $1.5$ TeV avoid too strong bounds on the MSSM part of model from $B\rightarrow X_s \gamma$ decays. In figure~\ref{fig:SceA}, the common areas of $b\rightarrow s\ell^+\ell^-$ explanations under other bounds show a larger value and region of $\lambda'^{\ast}_{223}\lambda'_{233}$ or $-\lambda'^{\ast}_{323}\lambda'_{333}$ for a heavier $m_{\tilde{L}'_2}$.
The values of $\lambda'^{\ast}_{323}\lambda'_{333}$ always have the negative sign compared with the positive $\lambda'^{\ast}_{223}\lambda'_{233}$, and their region sizes are nearly the same for the same value of $m_{\tilde{L}'_2}$.
The results in scenario A show that $b\rightarrow s\ell^+\ell^-$ anomalies in both $1\sigma$ and $2\sigma$ fits can be explained.
 
Combined with considering the $\Delta a_{\mu}$ deviation, we find the final common region to explain the $b\to s\ell^+\ell^-$ and $a_\mu$ anomalies simultaneously at $2\sigma$ level which are shown by the points on the left of $\Delta a_{\mu}$ bound line in figure~\ref{fig:SceA}(b). The result provides the $m_{\tilde{L}'_2}$ range as $370~{\rm GeV} \leqslant m_{\tilde{L}'_2} \leqslant 470$~GeV and edge values of $\lambda'$ combinations are collected in table~\ref{tab:SceA}. 
\begin{table}[t]
\centering
\setlength\tabcolsep{8pt}
\renewcommand{\arraystretch}{1.3}
\begin{tabular}{c|c|c}
        \hline
		$m_{\tilde{L}'_2}$~[GeV] & $\lambda'^{\ast}_{223}\lambda'_{233}$ & $\lambda'^{\ast}_{323}\lambda'_{333}$ \\
		\hline
        $370$ & $[0.14,0.30]$ & $[-0.26,-0.12]$ \\
        $420$ & $[0.17,0.37]$ & $[-0.32,-0.15]$ \\
        $470$ & $[0.20,0.44]$ & $[-0.38,-0.18]$ \\        
        \hline
	\end{tabular}
	\caption{The edge values of $\lambda'^{\ast}_{223}\lambda'_{233}$ and $\lambda'^{\ast}_{323}\lambda'_{333}$ related to different $m_{\tilde{L}'_2}$ for the simultaneous explanation of $b\to s\ell^+\ell^-$ and $a_\mu$ anomalies at $2\sigma$ level in scenario A.}
	\label{tab:SceA}
\end{table}

\subsubsection{Results in scenario B}
\begin{figure}[htbp]
	\centering
\includegraphics[width=1\textwidth]{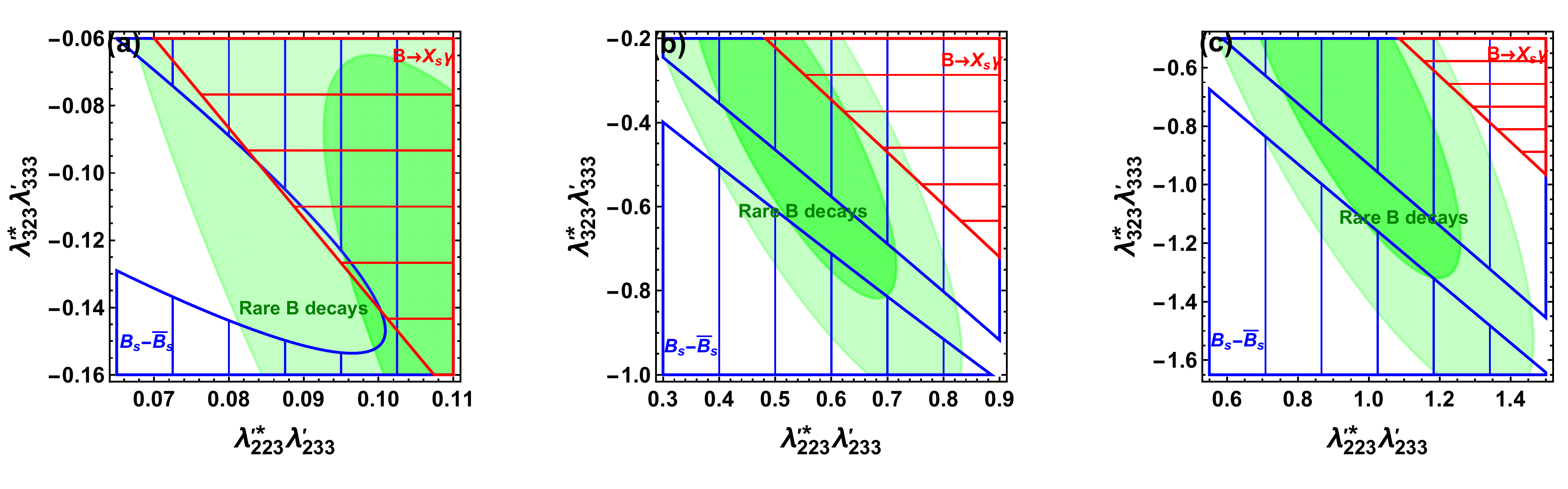}
	\caption{The regions of constraints in scenario B without $a_{\mu}$ data. The green regions are $1(2)\sigma$ favored ones with dark (light) opacity to satisfy the rare $B$-meson decay fits. At $2\sigma$ level, the hatched blue areas are excluded by $B_s-\bar{B}_s$ mixing and the hatched red areas are excluded by $B\rightarrow X_s \gamma$ decays. 
Besides, $m_{\tilde{L}'_3}=m_{\tilde{L}'_2}$ are fixed as $430$ (left), $750$ (middle) and $1000$~GeV (right).}\label{fig:SceB1}
\end{figure}

\begin{figure}[htbp]
	\centering
\includegraphics[width=0.95\textwidth]{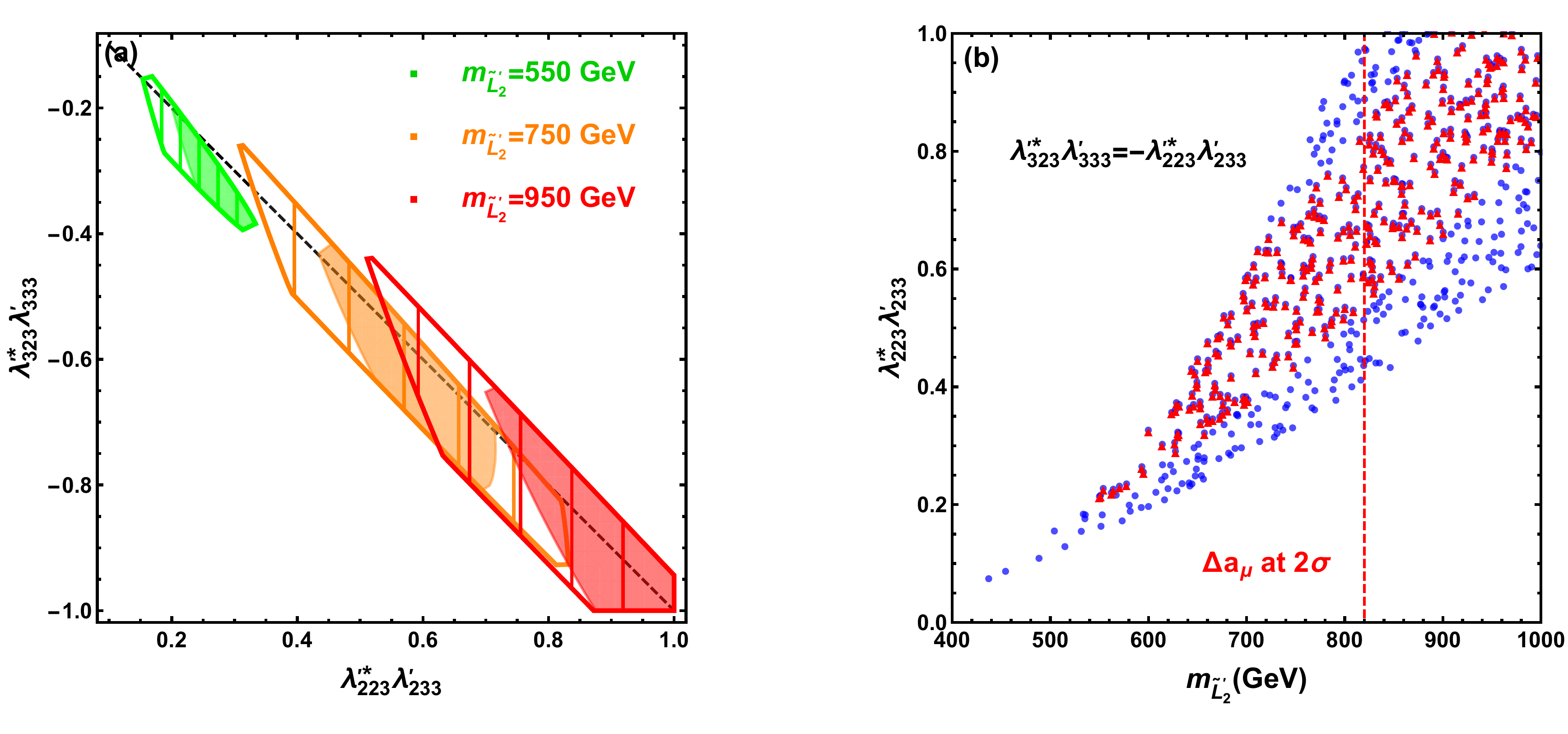}
	\caption{The common scopes of constraints in scenario B. Figure~\ref{fig:SceB2}(a) shows the common scopes constrained by the rare $B$-meson decay fits at $1\sigma$ level denoted by painted areas and $2\sigma$ level denoted by hatched areas, combined with other process constraints at $2\sigma$ level for $m_{\tilde{L}'_3}=m_{\tilde{L}'_2}=550$ (green), $750$ (orange) and $950$ GeV (red). Figure~\ref{fig:SceB2}(b) shows the regions which satisfy the rare $B$-meson decay fits with being $1(2)\sigma$ favored under the assumption $\lambda'^{\ast}_{323}\lambda'_{333}=-\lambda'^{\ast}_{223}\lambda'_{233}$, denoted by red (blue) points, with other process constraints being considered. The area on the left of the red dashed line is allowed to be accordant with $a_{\mu}$ data at $2\sigma$.}\label{fig:SceB2}
\end{figure}

In scenario B, the common scopes are shown at figure~\ref{fig:SceB1} and figure~\ref{fig:SceB2}.
Similar to scenario A, the values of allowed $\lambda'^{\ast}_{323}\lambda'_{333}$ are always negative compared with the positive $\lambda'^{\ast}_{223}\lambda'_{233}$ and the region sizes of their common scopes become larger as $m_{\tilde{L}'_2}$ varying heavier. 
As shown in figure~\ref{fig:SceB1}, 
there exist areas to explain $b\to s\ell^+\ell^-$ anomalies at $1\sigma$ level for rare $B$-meson decay fits, and the $B_s-\bar{B}_s$ mixing constrains mostly. While $B\rightarrow X_s \gamma$ decays provide no extra bounds when $m_{\tilde{L}'_2}\gtrsim 430$~GeV and even reaches TeV.  
When $m_{\tilde{L}'_2}$ blow around $430$~GeV, $1\sigma$ explanations will not be viable, and $B\rightarrow X_s \gamma$ decays provide extra bounds versus $B_s-\bar{B}_s$ mixing. 
In figure~\ref{fig:SceB2}(a), we compare the common region sizes for the different fixed $m_{\tilde{L}'_2}$ with each other and find that the deviation between the allowed region sizes of $\lambda'^{\ast}_{223}\lambda'_{233}$ and $-\lambda'^{\ast}_{323}\lambda'_{333}$ is small, up to around $0.1$ scale. Thus we further fix them equalling to each other and show $\lambda'^{\ast}_{223}\lambda'_{233}$ varying with increasing $m_{\tilde{L}'_2}$ only in figure~\ref{fig:SceB2}(b) and find that, the $1\sigma$ 
favored fit requires $m_{\tilde{L}'_2}\gtrsim550$~GeV and the region of $2\sigma$ fit has a broader size. When $m_{\tilde{L}'_2}$ below $550$ GeV, there also exist common regions of $1\sigma$ fit while in these regions, $\lambda'^{\ast}_{223}\lambda'_{233}+\lambda'^{\ast}_{323}\lambda'_{333}\neq 0$. 

Combined with $a_{\mu}$ data, the final common region of $400~{\rm GeV} \leqslant m_{\tilde{L}'_2} \leqslant 820$~GeV is required to explain the $b\to s\ell^+\ell^-$ and $a_\mu$ anomalies simultaneously at $2\sigma$ level and edge values of $\lambda'$ combinations are collected in table~\ref{tab:SceB}.

\begin{table}[bp]
\centering
\setlength\tabcolsep{8pt}
\renewcommand{\arraystretch}{1.3}
\begin{tabular}{c|c|c}
        \hline
		$m_{\tilde{L}'_2}$~[GeV] & $\lambda'^{\ast}_{223}\lambda'_{233}$ & $\lambda'^{\ast}_{323}\lambda'_{333}$ \\
		\hline
        $420$ & $[0.062,0.086]$ & $[-0.137,-0.063]$ \\
        $650$ & $[0.22,0.62]$ & $[-0.70,-0.17]$ \\
        $820$ & $[0.35,1.00]$ & $[-1.10,-0.30]$ \\        
        \hline
	\end{tabular}
	\caption{The same as table~\ref{tab:SceA} except for scenario B.}
	\label{tab:SceB}
\end{table}

\subsection{Predictions of $t\rightarrow c g$ decay}
As the numerical discussions above, we have the final parameter spaces of $m_{\tilde{L}'_2}$ as well as the coupling combinations $\lambda'^{\ast}_{223}\lambda'_{233}$ and $\lambda'^{\ast}_{323}\lambda'_{333}$ to explain related LFU violating anomalies. While these variables also provide NP effects on the top decays $t\rightarrow c V(h)$.

We have checked that our final parameter spaces can satisfy the most recent upper limits on the branching ratios of $t\rightarrow c V(h)$ decays at LHC easily. 
The NP contributions to the branching ratios of $t\rightarrow c V(h)$ depend on the term $\tilde{\lambda}'^{\ast}_{i2k}\tilde{\lambda}'_{i3k}f_{\tilde{l}_{Li}}$ where $f_{\tilde{l}_{Li}}$ stands for the loop integral including LH charged sleptons. And this term can be given by $\tilde{\lambda}'^{\ast}_{i2k}\tilde{\lambda}'_{i3k}f_{\tilde{l}_{Li}} \approx (\lambda'^{\ast}_{i2k}\lambda'_{i3k}+|\lambda'_{i2k}|^2 K_{cb}+|\lambda'_{i3k}|^2 K_{cb})f_{\tilde{l}_{Li}}$. Because of cancelling out in $\lambda'^{\ast}_{i2k}\lambda'_{i3k}f_{\tilde{l}_{Li}}$, the hierarchical structure between $\lambda'_{a2k}$ and $\lambda'_{a3k}$ ($a=2,3$) is considered to make prominent contributions and we set the large $\lambda'_{a3k}$ here. We keep restricting $k$ as $3$ and set $\lambda'_{233}=\lambda'_{333}=2$ or $3$.
\begin{figure}[htbp]
	\centering
\includegraphics[width=0.6\textwidth]{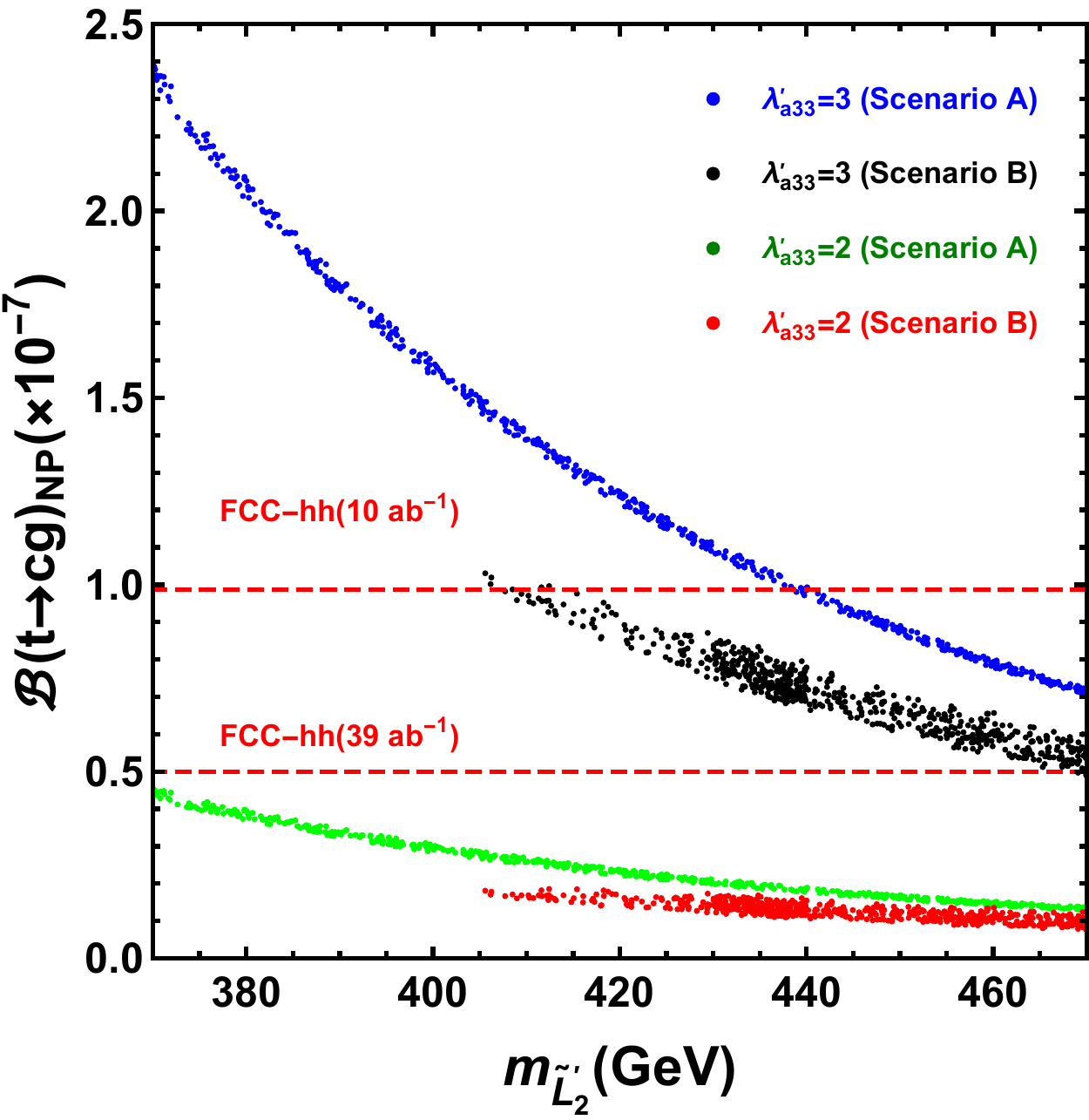}
	\caption{The predictions of ${\cal B}(t\rightarrow c g)_{\rm NP}$ compared with the prospect upper limit at $100$ TeV FCC-hh.}\label{fig:protcg}
\end{figure}

In the following we show that the prediction values of ${\cal B}(t\rightarrow c g)_{\rm NP}$ from the parameter spaces to explain $b\rightarrow s\ell^+\ell^-$ and $a_\mu$ anomalies can reach the sensitivity at the FCC-hh in figure~\ref{fig:protcg}.
One can see that in scenario A, when $370~{\rm GeV} \leqslant m_{\tilde{L}'_2} \leqslant 440$~GeV and $\lambda'_{a33}=3$, the prediction ${\cal B}(t\rightarrow c g)_{\rm NP}$ is higher than the prospect upper limit $9.87\times 10^{-8}$, at $100$ TeV FCC-hh for the integrated
luminosity of ${\cal L}=10$~${\rm ab}^{-1}$ of data through the triple-top signal~\cite{Khanpour:2019qnw}
, and the prediction in scenario B for the same $\lambda'_{a33}$ can also reach this upper limit.
When $\lambda'_{a33}$ is set to be $2$, the branching ratio is much lower and can not even reach the sensitivity at FCC-hh for the estimated ${\cal L}=39$~${\rm ab}^{-1}$ in both scenario A and B. We conclude that, at FCC-hh, this model signal on the $t\rightarrow cg$ transition has considerable possibilities to be found for sufficiently large $\lambda'_{a33}$, but the model can escape easily from the bound of this transition when the structure between $\lambda'_{a23}$ and $\lambda'_{a33}$ is not hierarchical enough.

%%%%%%%%%%%%%%%%%%%%%%%%%%%%%%%%%%%%%%%%%%%%%%%%%%%%%%
\section{Conclusions}\label{sec:conclusion}
Recent measurements on the transition $b\rightarrow s\ell^+\ell^-$ reveal the deviations from SM predictions. The most motivative $R_K^{(\ast)}$ anomaly and anomalies from other observables like $P'_5$, called $b\rightarrow s\ell^+\ell^-$ anomalies collectively, suggest the NP of LFU violation may exist. Besides, this NP may also affect the enduring muon anomalous magnetic moment, $(g-2)_\mu$ problem.

In this work, we have studied the chiral mixing effects of sneutrinos in the $R$-parity violating MSSM with inverse seesaw mechanisms to explore the explanation of $b\rightarrow s\ell^+\ell^-$ anomalies with $(g-2)_\mu$ problem simultaneously. Here all the one-loop contributions to $b\rightarrow s\ell^+\ell^-$ processes are scrutinized under the assumption of a single value $k$. Among them, the contributions of chiral mixing between LH and singlet (s)neutrinos within superpotential term $\lambda'_{ijk} \hat L_i \hat Q_j \hat D_k$ are given for the first time to our knowledge. To explain $b\rightarrow s\ell^+\ell^-$ anomalies in this model, two kinds of model-independent global fits are adopted. One is the single-parameter scenario of $C^{\rm NP}_{9,\mu}=-C^{\rm NP}_{10,\mu}$ and the other scenario is the double-parameter one that $(\pm)C_{\rm V}$ contributes to the $C^{\rm NP}_{9(10),\mu}$ part in 
$\mu^+\mu^-$ channel and $C_{\rm U}$ contributes to $C^{\rm NP}_{9}$ part in both $\mu^+\mu^-$ and $e^+e^-$ channels. Then in the numerical analyses, we find that $b\rightarrow s\ell^+\ell^-$ and $(g-2)_\mu$  anomalies can be explained simultaneously in both scenario A and B. The main constraints among related processes are from $B_s-\bar{B}_s$ mixing covering $B\rightarrow X_s\gamma$ decay mostly but the other tree-level and one-loop processes provide no effective bounds. At last we make a prospect that NP contributions to $t\rightarrow cg$ process can reach the sensitivity at FCC-hh in parts of the parameter spaces of this model.  

\section*{Acknowledgements}
We thank Yi-Lei Tang and Chengfeng Cai for valuable discussions. 
This work is supported in part by the National Natural Science Foundation of China under Grant No.~11875327, the Fundamental Research
Funds for the Central Universities, and the Sun Yat-Sen University Science Foundation.

\appendix
\section{\texorpdfstring{One-loop box contributions in RPV-MSSMIS}{}}\label{app:bsllCbox}
In this appendix, we list the whole Wilson coefficients from the one-loop box diagrams of $b\rightarrow s\ell^+\ell^-$ in RPV-MSSMIS without the extra assumption of a single value $k$.

The LH-quark-current contributions of chargino box diagrams to $b\rightarrow s\ell^+\ell^-$ process are given by
\begin{align}\label{eq:chaWcs}
C^{\chi^\pm}_{9,\ell}=&-C^{\chi^\pm}_{10,\ell}=-\frac{\sqrt{2}\pi^2 i}{2G_F \eta_t e^2} \Bigl(
g^2_2 K_{i3} K^{\ast}_{i2} V^{\ast}_{m1}V_{n1}
(g_2V_{m1}\tilde{\cal V}^{\cal I}_{v\ell}-
V_{m2}Y^{\cal I}_{\ell v}) \notag\\
&(g_2V^{\ast}_{n1}\tilde{\cal V}^{\cal I}_{v\ell}-
V^{\ast}_{n2}Y^{\cal I}_{\ell v}) 
D_2[m_{\tilde{\nu}^{\cal I}_{v}},m_{\chi^\pm_m},m_{\chi^\pm_n},m_{\tilde{u}_{Li}}] \notag\\
&+y^2_{u_i} K_{i3}K^{\ast}_{i2} V^{\ast}_{m2}V_{n2}
(g_2V_{m1}\tilde{\cal V}^{\cal I}_{v\ell}-
V_{m2}Y^{\cal I}_{\ell v}) \notag\\
&(g_2V^{\ast}_{n1}\tilde{\cal V}^{\cal I}_{v\ell}-
V^{\ast}_{n2}Y^{\cal I}_{\ell v}) 
D_2[m_{\tilde{\nu}^{\cal I}_{v}},m_{\chi^\pm_m},m_{\chi^\pm_n},m_{\tilde{u}_{Ri}}] \notag\\
&-\lambda'^{\cal I}_{v3k} \lambda'^{\cal I \ast}_{v'2k} (g_2 V^{\ast}_{m1} \tilde{\cal V}^{\cal I}_{v\ell}-V^{\ast}_{m2} Y^{\cal I}_{\ell v})
(g_2 V_{m1} \tilde{\cal V}^{\cal I}_{v'\ell}-V_{m2} Y^{\cal I}_{\ell v'}) D_2[m_{\tilde{\nu}^{\cal I}_{v}},m_{\tilde{\nu}^{\cal I}_{v'}},m_{\chi^\pm_m},m_{d_k}] \notag\\
&-\tilde{\lambda}'_{\ell ik} \tilde{\lambda}'^{\ast}_{\ell jk} g^2_2 K_{i3}K^{\ast}_{j2} |V_{m1}|^2 D_2[m_{\tilde{u}_{Li}},m_{\tilde{u}_{Lj}},m_{\chi^\pm_m},m_{d_k}] \notag\\
&+\tilde{\lambda}'_{\ell ik} \lambda'^{\cal I \ast}_{v2k} (g_2 K_{i3} V^{\ast}_{m1})(g_2 V_{m1} \tilde{\cal V}^{\cal I}_{v\ell}-V_{m2} Y^{\cal I}_{\ell v}) D_2[m_{\tilde{\nu}^{\cal I}_{v}},m_{\tilde{u}_{Li}},m_{\chi^\pm_m},m_{d_k}] \notag\\
&+\tilde{\lambda}'^{\ast}_{\ell ik} \lambda'^{\cal I}_{v3k} (g_2 K^{\ast}_{i2} V_{m1})(g_2 V^{\ast}_{m1} \tilde{\cal V}^{\cal I}_{v\ell}-V^{\ast}_{m2} Y^{\cal I}_{\ell v}) D_2[m_{\tilde{\nu}^{\cal I}_{v}},m_{\tilde{u}_{Li}},m_{\chi^\pm_m},m_{d_k}]
\Bigr),
\end{align}
where the Yukawa couplings $y_{u_i}={\sqrt{2}m_{u_i}}/{v_u}$ and $Y^{\cal I}_{\ell v}={(Y_\nu)}_{j\ell} \tilde{\cal V}^{\cal I \ast}_{v(j+3)}$. While the corresponding RH-quark-current contributions are
\begin{align}\label{eq:chaWcsRH}
C^{\prime\chi^\pm}_{9,\ell}=&-C^{\prime\chi^\pm}_{10,\ell}=-\frac{\sqrt{2}\pi^2 i}{2G_F \eta_t e^2}
\lambda'^{\cal I}_{vi2} \lambda'^{\cal I \ast}_{v'i3} (g_2 V^{\ast}_{m1} \tilde{\cal V}^{\cal I}_{v\ell}-V^{\ast}_{m2} Y^{\cal I}_{\ell v}) \notag\\
&(g_2 V_{m1} \tilde{\cal V}^{\cal I}_{v'\ell}-V_{m2} Y^{\cal I}_{\ell v'}) D_2[m_{\tilde{\nu}^{\cal I}_{v}},m_{\tilde{\nu}^{\cal I}_{v'}},m_{\chi^\pm_m},m_{d_i}].
\end{align}

The contributions of $W/H^\pm$ box diagrams to $b\rightarrow s\ell^+\ell^-$ process are given by
\begin{align}\label{eq:WHWcs}
C^{W/H^\pm}_{9,\ell}=&-C^{W/H^\pm}_{10,\ell}=-\frac{\sqrt{2}\pi^2 i}{2G_F \eta_t e^2} \Bigl(
y^2_{u_i} K_{i3}K^{\ast}_{i2} Z^2_{H_{h2}}Z^2_{H_{h'2}} |Y^{\cal N}_{\ell v}|^2 D_2[m_{\nu_v},m_{u_i},m_{H_h},m_{H_{h'}}] \notag\\
&-4g^2_2 m_{u_i}y_{u_i} m_{\nu_v} K_{i3}K^{\ast}_{i2} Z^2_{H_{h2}} {\text{Re}}({\cal V}_{v\ell} Y^{\cal N \ast}_{\ell v}) D_0[m_{\nu_v},m_{u_i},m_W,m_{H_{h}}] \notag\\
&+5g^4_2 K_{i3}K^{\ast}_{i2} |{\cal V}_{v\ell}|^2 D_2[m_{\nu_v},m_{u_i},m_W,m_W] \notag\\
&+Z^2_{H_{h2}} Y^{\cal N \ast}_{\ell v} Y^{\cal N}_{\ell v'} \lambda'^{\cal N}_{v3k} \lambda'^{\cal N \ast}_{v'2k} D_2[m_{\nu_v},m_{\nu_{v'}},m_{H_h},m_{\tilde{d}_{Rk}}] \notag\\
&-2g^2_2 m_{\nu_v} m_{\nu_{v'}} {\cal V}^{\ast}_{v\ell} {\cal V}_{v'\ell} \lambda'^{\cal N}_{v3k} \lambda'^{\cal N \ast}_{v'2k} D_0[m_{\nu_v},m_{\nu_{v'}},m_W,m_{\tilde{d}_{Rk}}] \notag\\
&+2 m_{\nu_v}m_{\nu_{v'}} Z^2_{H_{h2}} Y^{\cal N}_{\ell v} Y^{\cal N \ast}_{\ell v'} 
\lambda'^{\cal N}_{v3k} \lambda'^{\cal N \ast}_{v'2k}
D_0[m_{\nu_v},m_{\nu_{v'}},m_{H_h},m_{\tilde{d}_{Rk}}] \notag\\
&+2 m_{u_i}m_{u_j} y_{u_i}y_{u_j} K_{i3}K^{\ast}_{j2} Z^2_{H_{h2}}
\tilde{\lambda}'_{\ell ik} \tilde{\lambda}'^{\ast}_{\ell jk}
D_0[m_{u_i},m_{u_j},m_{H_h},m_{\tilde{d}_{Rk}}] \notag\\
&-g^2_2 {\cal V}_{v\ell} {\cal V}^{\ast}_{v'\ell} \lambda'^{\cal N}_{v3k} \lambda'^{\cal N \ast}_{v'2k}
D_2[m_{\nu_v},m_{\nu_{v'}},m_{W},m_{\tilde{d}_{Rk}}] \notag\\
&-g^2_2 K_{i3}K^{\ast}_{j2} \tilde{\lambda}'_{\ell ik} \tilde{\lambda}'^{\ast}_{\ell jk}
D_2[m_{u_i},m_{u_j},m_{W},m_{\tilde{d}_{Rk}}] \notag\\
&-2 m_{u_i}y_{u_i} m_{\nu_v} K_{i3} Z^2_{H_{h2}} Y^{\cal N \ast}_{\ell v} \tilde{\lambda}'_{\ell ik} 
\lambda'^{\cal N \ast}_{v2k}
D_0[m_{u_i},m_{\nu_v},m_{H_h},m_{\tilde{d}_{Rk}}] \notag\\
&-2 m_{u_i}y_{u_i} m_{\nu_v} K^{\ast}_{i2} Z^2_{H_{h2}} Y^{\cal N}_{\ell v} \tilde{\lambda}'^{\ast}_{\ell ik} 
\lambda'^{\cal N}_{v3k}
D_0[m_{u_i},m_{\nu_v},m_{H_h},m_{\tilde{d}_{Rk}}] \notag\\ 
&+g^2_2 K^{\ast}_{i2} {\cal V}_{v\ell} \tilde{\lambda}'^{\ast}_{\ell ik} 
\lambda'^{\cal N}_{v3k}
D_2[m_{u_i},m_{\nu_v},m_W,m_{\tilde{d}_{Rk}}] \notag\\
&+g^2_2 K_{i3} {\cal V}^{\ast}_{v\ell} \tilde{\lambda}'_{\ell ik} 
\lambda'^{\cal N \ast}_{v2k}
D_2[m_{u_i},m_{\nu_v},m_W,m_{\tilde{d}_{Rk}}]
\Bigr), \\
\ \notag\\
C^{\prime W/H^\pm}_{9,\ell}=&-C^{\prime W/H^\pm}_{10,\ell}=-\frac{\sqrt{2}\pi^2 i}{2G_F \eta_t e^2} \Bigl( 
-2 Z^2_{H_{h2}} Y^{\cal N\ast}_{\ell v} Y^{\cal N}_{\ell v'}
\lambda'^{\cal N}_{v'i2} \lambda'^{\cal N \ast}_{vi3} 
m_{\nu_v} m_{\nu_{v'}} 
D_0[m_{\nu_v},m_{\nu_{v'}},m_{H_h},m_{\tilde{d}_{Li}}] \notag\\
&- Z^2_{H_{h2}} Y^{\cal N}_{\ell v} Y^{\cal N\ast}_{\ell v'}
\lambda'^{\cal N}_{v'i2} \lambda'^{\cal N \ast}_{vi3}
D_2[m_{\nu_v},m_{\nu_{v'}},m_{H_h},m_{\tilde{d}_{Li}}] \notag\\
&+g^2_2 {\cal V}_{v\ell} {\cal V}^{\ast}_{v'\ell} 
\lambda'^{\cal N}_{vi2} \lambda'^{\cal N \ast}_{v'i3}
D_2[m_{\nu_v},m_{\nu_{v'}},m_W,m_{\tilde{d}_{Li}}] \notag\\
&+2 g^2_2 {\cal V}_{v\ell} {\cal V}^{\ast}_{v'\ell} 
\lambda'^{\cal N}_{v'i2} \lambda'^{\cal N \ast}_{vi3}
m_{\nu_v} m_{\nu_{v'}} 
D_0[m_{\nu_v},m_{\nu_{v'}},m_W,m_{\tilde{d}_{Li}}]
\Bigr),
\end{align}
where the mixing matrix elements $Z_{H_{12}}=-\sin \beta$, $Z_{H_{22}}=-\cos\beta$ with Goldstone mass $m_{H_1}=m_W$ and charged Higgs mass  $m_{H_2}=m_{H^\pm}$ and $Y^{\cal N}_{\ell v}={(Y_\nu)}_{j\ell} {\cal V}^{\ast}_{v(j+3)}$.

The contributions of $4\lambda'$ box diagrams to $b\rightarrow s\ell^+\ell^-$ process are given by
\begin{align}
C^{4\lambda'}_{9,\ell}=&-C^{4\lambda'}_{10,\ell}=-\frac{\sqrt{2}\pi^2 i}{2G_F \eta_t e^2} \Bigl(
\tilde{\lambda}'_{\ell ik}  \tilde{\lambda}'^{\ast}_{\ell ik'}
\lambda'^{\cal N}_{v3k'} 
\lambda'^{\cal N \ast}_{v2k}
D_2[m_{\nu_v},m_{u_i},m_{\tilde{d}_{Rk}},m_{\tilde{d}_{Rk'}}] \notag\\
&+\tilde{\lambda}'_{\ell ik'}  \tilde{\lambda}'^{\ast}_{\ell ik}
\lambda'^{\cal I}_{v3k} 
\lambda'^{\cal I \ast}_{v2k'}
D_2[m_{\tilde{\nu}^{\cal I}_v},m_{\tilde{u}_{Li}},m_{d_k},m_{d_{k'}}]
\Bigr),\\
\ \notag\\
C^{\prime4\lambda'}_{9,\ell}=&-C^{\prime4\lambda'}_{10,\ell}=-\frac{\sqrt{2}\pi^2 i}{2G_F \eta_t e^2} \Bigl(
\tilde{\lambda}'_{\ell ij}  \tilde{\lambda}'^{\ast}_{\ell ij'}
\lambda'^{\cal N}_{vj2} 
\lambda'^{\cal N \ast}_{vj'3}
D_2[m_{\nu_v},m_{u_i},m_{\tilde{d}_{Lj}},m_{\tilde{d}_{Lj'}}] \notag\\
&-\tilde{\lambda}'_{\ell j'k} \tilde{\lambda}'^{\ast}_{\ell jk}
\tilde{\lambda}'_{ij2} \tilde{\lambda}'^{\ast}_{ij'3}
\bigl(D_2[m_{l_i},m_{\tilde{u}_{Lj}},m_{\tilde{u}_{Lj'}},m_{d_k}]
+D_2[m_{\tilde{l}_{Li}},m_{u_j},m_{u_{j'}},m_{\tilde{d}_{Rk}}] \bigr) \Bigr).
\end{align}

The contributions of neutralino box diagrams only contain RH-quark-current parts, which are given by
\begin{align}\label{eq:neuWcs}
C^{\prime \chi^0}_{9,\ell}=&-C^{\prime \chi^0}_{10,\ell}=-\frac{\sqrt{2}\pi^2 i}{2G_F \eta_t e^2} \Bigl(
\frac{1}{2} (g_1 {N_{n1}} + g_2 {N_{n2}})^2
\tilde{\lambda}'_{\ell i2} \tilde{\lambda}'^{\ast}_{\ell i3}
D_2[m_{\tilde{l}_{L\ell}},m_{\tilde{l}_{L\ell}},m_{u_i},m_{\chi^0_n}] \notag\\
&+\frac{2}{9} g^2_1 |{N_{n1}}|^2 \tilde{\lambda}'_{\ell i2} \tilde{\lambda}'^{\ast}_{\ell i3}
D_2[m_{u_i},m_{\chi^0_n},m_{\tilde{s}_{R}},m_{\tilde{b}_{R}}] \notag\\
&-\frac{1}{3} {N_{n1}} g_1 (g_1 {N_{n1}} + g_2 {N_{n2}})
\tilde{\lambda}'_{\ell i2} \tilde{\lambda}'^{\ast}_{\ell i3}
\bigl(D_2[m_{\tilde{l}_{L\ell}},m_{u_i},m_{\chi^0_n},m_{\tilde{s}_{R}}]+
D_2[m_{\tilde{l}_{L\ell}},m_{u_i},m_{\chi^0_n},m_{\tilde{b}_{R}}] \bigr)
\Bigr).
\end{align}
 
\bibliographystyle{JHEP}
%\bibliography{ref,ref_prd,references}
\bibliography{ref}

\end{document}